# THE ASTROPHYSICAL JOURNAL

Greater climate sensitivity and variability on TRAPPIST-1e than Earth

Running head: TRAPPIST-1e climate sensitivity

Assaf Hochman[1,2,3], Paolo De Luca[4], Thaddeus D. Komacek[5]

**Abstract**

The atmospheres of rocky exoplanets are close to being characterized by astronomical observations, in part due to the commissioning of the James Webb Space Telescope. These observations compel us to understand exoplanetary atmospheres, in the voyage to find habitable planets. With this aim, we investigate the effect that $CO_2$ partial pressure ($pCO_2$) has on exoplanets' climate variability, by analyzing results from ExoCAM model simulations of the tidally locked TRAPPIST-1e exoplanet, an Earth-like aqua-planet and Earth itself. First, we relate the differences between the planets to their elementary parameters. Then, we compare the sensitivity of the Earth analogue and TRAPPIST-1e's surface temperature and precipitation to $pCO_2$. Our simulations suggest that the climatology and extremes of TRAPPIST-1e's temperature are ~1.5 times more sensitive to $pCO_2$ relative to Earth. The precipitation sensitivity strongly depends on the specific region analyzed. Indeed, the precipitation near mid-latitude and equatorial sub-stellar regions of TRAPPIST-1e is more sensitive to $pCO_2$, and the precipitation sensitivity is ~2 times larger in TRAPPIST-1e. A dynamical systems perspective, which provides information about how the atmosphere evolves in phase-space, provides additional insights. Notably, an increase in $pCO_2$, results in an increase in atmospheric persistence on both planets, and the persistence of TRAPPIST-1e is more sensitive to $pCO_2$ than Earth. We conclude that the climate of TRAPPIST-1e may be more sensitive to $pCO_2$, particularly on its dayside. This study documents a new pathway for understanding the effect that varying planetary parameters have on the climate variability of potentially habitable exoplanets and on Earth.

[1] Corresponding author assaf.hochman@kit.edu Assaf.Hochman@mail.huji.ac.il
[2] The Fredy and Nadine Hermann Institute of Earth Sciences, The Hebrew University of Jerusalem (HUJI), Edmond J. Safra Campus, Givat Ram, 9190401 Jerusalem, Israel.
[3] Institute of Meteorology and Climate Research, Department of Tropospheric Research, (IMK-TRO), Karlsruhe Institute of Technology (KIT), 76344 Eggenstein-Leopoldshafen, Germany.
[4] Barcelona Supercomputing Center (BSC), 08034 Barcelona, Spain.
[5] Department of Astronomy, University of Maryland, 4296 College Park, Maryland, USA.

**1.INTRODUCTION**

The astronomical study of exoplanet climates is approaching the point in which climate variability will need to be taken into account to characterize exoplanet atmospheres (Rauscher , et al., 2007; Dobbs-Dixon, et al., 2010; Komacek & Showman, 2020; Charnay, et al., 2021; May, et al., 2021). The processes that drive climate variability on exoplanets are a priori unknown, but for rocky exoplanets, they are likely to include weather and climate extreme events similar to Earth, such as heatwaves, hurricanes, drought and cold spells (Stott, et al., 2004; Emanuel, 1988; Mukherjee, et al., 2018; de vries, et al., 2012). Climate variability may also affect habitability by inducing strong time-variations in the fractional extent of a planetary surface that has habitable conditions (Colose, et al., 2019; Del Genio, et al., 2019; Jansen, et al., 2019). In this work, we generalize the concept of climate extremes to consider how these shape the climates of exoplanets. We define such planetary atmospheric extremes as extreme climate events in exoplanet atmospheres that have consequences for either local habitability or observable properties, including climate variability (Komacek, et al., 2020; Yan & Yang , 2020). Given the broad possible parameter space of exoplanet atmospheres, studying exoplanetary climate dynamics and extremes can place Earth's climate variability on the continuum of planetary climate states in the Milky Way.

Climate extremes are of important relevance to society on Earth due to their harmful impacts as emphasized by the recent Intergovernmental Panel on Climate Change report (IPCC, 2021). Correspondingly, there is a large body of scientific literature focusing on how changes in greenhouse gas concentrations, particularly $CO_2$, may influence climate dynamics and extremes on Earth (e.g., Sillmann, et al., 2013; Schewe, et al., 2019; Vogel, et al., 2019). Such studies use either direct observations (Easterling, et al., 2016), indirect past climate reconstructions (PAGES2k consortium, 2017) and/or general circulation model (GCM) simulations (Ajjur & Al-Ghamdi, 2021). With respect to the latter, a very important contribution to the study of climate extremes and dynamics on Earth are the different phases of the Coupled Model Inter-comparison Project (CMIP3, CMIP5 and CMIP6; Taylor, et al., 2012; Meehl, et al., 2007; Eyring, et al., 2016) and the Coordinated Downscaling Experiment (CORDEX; Giorgi, et al., 2009). From these overarching programs, it can generally be concluded that climate extremes, particularly temperature and precipitation extremes, are strongly influenced by variations in $CO_2$ concentrations on Earth and that these are, and will continue to increase in the future, in both frequency and intensity (IPCC, 2021). Indeed, the recent CMIP6 GCMs show an even stronger climate sensitivity to $CO_2$ doubling than suggested in earlier experiments. This rather surprising finding most probably relates to the better representation of clouds in the new versions of the GCMs (Zelinka, et al., 2020).

Studies of exoplanet atmospheres are useful to place Earth in a broader context, especially by considering how varying planetary parameters away from that of Earth impact atmospheric circulation and planetary climate. Recent studies based on GCM simulations of exoplanet atmospheres have provided insight into how atmospheric dynamics depend on key planetary properties such as instellation, rotation rate and planetary radius/gravity (Kaspi & Showman, 2015; Way, et al., 2018; Komacek & Abbot, 2019). In tandem, an assortment of studies have demonstrated that there is also a large impact of planetary parameters on the climate dynamics of tidally locked rocky exoplanets orbiting late-type dwarf stars (Noda, et al., 2017; Haqq-Misra, et al., 2018; Komacek & Abbot, 2019; Yang, et al., 2019). Notably, Haqq-Misra et al. (2018) proposed that the atmospheric dynamics of tidally locked rocky exoplanets could be classified into rotation regimes by the combination of two dynamical length scales: the Rossby deformation radius and Rhines scale. In general, recently developed frameworks to understand the impact of planetary parameters on atmospheric circulation and climate have demonstrated that each exoplanet must be considered individually to make deterministic predictions of its climate state and observable properties.

Though there have been a wide range of studies on the dependence of the mean climate of temperate tidally locked exoplanets orbiting single stars on their planetary properties, there have been limited studies on their climate dynamics and variability. Recently, the climate variability of sub-Neptune K2-18b and the temperate terrestrial exoplanet TRAPPIST-1e have been studied (Charnay, et al., 2021; May, et al., 2021). Additionally, exploration of climate extremes on tidally locked rocky exoplanets has been focused on tropical cyclones, which have been demonstrated to potentially occur on a wide range of rocky exoplanets (Yan & Yang , 2020). One type of exoplanets in which emergent climate variability is expected are circumbinary planets, which present a natural case study of climate extremes due to the time-varying irradiation from the host binary star system. Recent 1D and 3D studies of the climate of temperate circumbinary exoplanets have found circumbinary-induced climate variations of up to ~10°K, which may result in climate extremes that influence planetary habitability (Haqq-Misra, et al., 2019; Wolf, et al., 2021).

In this work, we focus on the temperate rocky exoplanet TRAPPIST-1e, to study its planetary climate extremes and dynamics and compare them to both modern Earth and to Earth-like exoplanets. We concentrate on TRAPPIST-1e as a test case for two reasons. First, TRAPPIST-1e is well-studied by a range of previous exoplanet GCM experiments, and constitutes a benchmark planet for model development (Wolf, 2017; Turbet, et al., 2018; Fauchez, et al., 2019; Fauchez, et al., 2021; May, et al., 2021; Sergeev, et al., 2021; Turbet, et al., 2021). Second, TRAPPIST-1e will be a prime temperate exoplanet for atmospheric characterization with the James Webb Space Telescope, which may potentially enable the detection of an atmosphere along with key atmospheric species and bio-signature pairs such as carbon dioxide and methane (Krissansen-Totton, et al., 2018; Fauchez, et al., 2019; Lustig-Yaeger, et al., 2019; Mikal-Evans, 2022). Here, we conduct novel long-timescale GCM integrations of the atmospheric circulation of TRAPPIST-1e with varying $CO_2$ partial pressure ($pCO_2$) along with a comparison set of Earth-like exoplanet model simulations. We then study the climate extremes and dynamics in both simulation sets, in consort with recent Earth reanalysis data, and compare them to determine how climate extremes and dynamics differ between planets like TRAPPIST-1e, that orbit close-in to late-type M dwarf stars, Earth-like exoplanets orbiting Sun-like stars, and Earth itself.

In this study, we characterize the time series dynamics of a planet's atmosphere by leveraging recent developments in dynamical systems theory. These advancements let us define instantaneous atmospheric patterns in terms of the persistence ($\theta^{-1}$), which provides indications about the mean residence time of recurrences around the state of interest in the phase space, and local dimension ($d$), which advises on how the atmosphere revolves around a state of interest (Faranda, et al., 2017). These metrics are intuitively related to how the atmosphere evolves with time. A highly persistent (low $\theta$), low dimensional (low $d$), state will change less in time than a low-persistence (high $\theta$), high dimensional (high $d$) one (Messori, et al., 2017). This theoretical approach has been found very useful to study atmospheric variability in different regions on Earth (Hochman, et al., 2019; Hochman, et al., 2020; Hochman, et al., 2022; Hochman, et al., 2021; Hochman , et al., 2021; De Luca, et al., 2020a; De Luca, et al., 2020b). However, this novel perspective has not yet been employed at the planetary scale for neither Earth nor exoplanet atmospheres.

This manuscript is organized as follows: we describe the present-day Earth data in Section 2. Model simulations conducted and data analysis methods are described in Section 3. We portray our results individually for the spatial difference between the climate extremes of TRAPPIST-1e and Earth-like exoplanets (Sect. 4.1). Then we present spatially averaged climate extremes (Sect. 4.2), and climate dynamics (Sect. 4.3) sensitivity to pCO$_2$ of TRAPPIST-1e and Earth-like exoplanets. We discuss the implications of our results for the observational characterization of exoplanet climates and delineate conclusions in Section 5.

## 2.DATA

As a representation of present-day Earth, we use the very recent version of the European Centre for Medium-Range Weather Forecasts (ECMWF) ERA5 reanalysis over 1979-2020, with a horizontal grid spacing of 0.25° × 0.25° (Hersbach, et al., 2020). From this reanalysis product, we extract daily maximum 2m temperature (*Tmax*, °C), minimum 2m temperature (*Tmin*, °C; see Appendix) and daily total precipitation (*Prec*, mm d$^{-1}$). Then, we re-gridded the data from longitude-latitude 1440x720 to longitude-latitude 72x46 in order to make it comparable with the ExoCAM simulations of TRAPPIST-1e and the Earth-like aqua-planet.

## 3.METHODS

### 3.1 ExoCAM MODEL SIMULATIONS

To study climate extremes and dynamics of TRAPPIST-1e and compare it those of an Earth-analogue planet, we utilize the ExoCAM GCM (publicly available at: https://github.com/storyofthewolf/ExoCAM ; Wolf, et al., 2022). ExoCAM is a well-established model that has been previously utilized to study the atmospheric circulation of a broad range of exoplanets (Kopparapu, et al., 2017; Wolf, 2017; Haqq-Misra, et al., 2018; Komacek & Abbot, 2019; Yang, et al., 2019; Suissa, et al., 2020; May, et al., 2021). The model is built from the Community Atmosphere Model (CAM) version 4 (Neale, et al., 2010), and includes routines to consider planetary properties and orbital configurations of a range of exoplanets, along with the novel non-gray correlated-k radiative transfer scheme ExoRT (https://github.com/storyofthewolf/ExoRT).

In this work, we conduct a suite of atmospheric model simulations of both TRAPPIST-1e and an Earth-analogue planet. To simulate the atmosphere of TRAPPIST-1e, we initialize the atmosphere from the end-state of previous simulations of TRAPPIST-1e (May, et al., 2021). Specifically, we conduct a set of three simulations with 1 bar of $N_2$ and varying $pCO_2$ from $10^{-2}$ - 1 bar with intervals of an order-of-magnitude between each level of $pCO_2$ (Referred to from this point on as 'Low' 'Mid' and 'High' $pCO_2$ scenarios, respectively). This sweep of $pCO_2$ was chosen to cover the range of possible climate states for TRAPPIST-1e (Wolf, 2017). In all three simulations, we use a radius of TRAPPIST-1e of 0.92 Earth radii, surface gravity of 9.12 m s$^{-2}$, and incident stellar flux of 900.85 W m$^{-2}$ with an incident stellar spectrum corresponding to an M-dwarf with effective temperature of 2600 K (Allard, et al., 2007). We assume that TRAPPIST-1e is tidally locked with an orbital and rotation period of 6.10 Earth days that its orbit has zero eccentricity, and that TRAPPIST-1e has zero obliquity. We consider the surface of the planet to be a global ocean (i.e., an aqua-planet) with a depth of 50 m. We include a thermodynamic sea ice scheme (Bitz, et al., 2012), but do not include ocean heat transport. From the initial condition of May et al. (2021), we then continue each simulation for 80 years to obtain daily mean output that we use to analyze climate extremes and dynamics.

To compare with the TRAPPIST-1e simulations described above, we conduct a similar suite of model simulations with planetary properties comparable to that of Earth. We likewise conduct three simulations covering $pCO_2$ ranging from $10^{-2}$ - 1 bar, assuming 1 bar of background $N_2$. To enable direct comparison with the simulations of TRAPPIST-1e, these simulations also assume an aqua-planet surface with a 50 m deep slab ocean and zero ocean heat transport, along with zero orbital eccentricity and zero planetary obliquity. These Earth-analogue simulations use a planetary radius of 6.37122 × $10^6$ m, surface gravity of 9.80616 m s$^{-2}$, rotation period of 8.64×$10^4$ s, and an incident stellar flux of 1360 W m$^{-2}$ with an incident stellar spectrum corresponding to our Sun.

All simulations presented in this work, both for the TRAPPIST-1e and Earth-analogue cases, have 40 vertical levels and a horizontal grid spacing of 4° × 5° (or longitude-latitude 72x46). The dynamical time step of the simulations is 30 minutes, and the radiative time step is 90 minutes. Though each ExoCAM simulation includes ~45 years of initial spin-up, we only analyze the final 80 years of daily output from all cases.

## 3.2 COMPUTATION OF CLIMATE EXTREMES

TRAPPIST-1e *Tmax* and precipitation mean climatology for each $pCO_2$ scenario (Fig. 1) are computed by averaging over their entire period (i.e., 80 years). We define extremes at the grid-box level, by retaining daily atmospheric values that exceed the 95th percentile of their distribution. We further consider six sub-regions, namely: mid-latitude anti-stellar (30 - 60N, 345 - 15E), mid-latitude sub-stellar (30 - 60N, 165 -195E), equatorial anti-stellar (-15 - 15N, 345 - 15E), equatorial sub-stellar (-15 - 15N, 165 - 195E), equatorial west-terminator (-15 - 15N, 75 - 105E) and equatorial east-terminator (-15 - 15N, 255 - 285E; Fig. 1 a). We choose these sub-regions since they separate the dayside and nightside hemispheres along with the terminator regions in order to distinguish the impacts of tidal locking on climate variability.

The spatial patterns of TRAPPIST-1e and the Earth-analogue extremes are computed as follows: i) for each $pCO_2$ simulation and grid-box take the extremes (i.e. values > 95th percentile); ii) for each grid-box take the mean of extremes. For ERA5 we use the same procedure. Then, the spatial differences ($\Delta$) between Earth and TRAPPIST-1e, ERA5 and TRAPPIST-1e, and ERA5 and Earth are computed by subtracting the medians of the extremes. We further compute, at the grid-box level, the mean anomalies of climate extremes for TRAPPIST-1e, Earth and ERA5. The mean anomalies of extremes for TRAPPIST-1e, Earth and ERA5 are computed based on ERA5 (1979-2020) daily climatology and are displayed in boxplots.

## 3.3 DYNAMICAL SYSTEMS METRICS

To characterize the climate dynamics of TRAPPIST-1e, the Earth-analogue and ERA5 reanalysis data sets we use a recently developed approach, which combines Poincaré recurrences with extreme value theory (Lucarini, et al., 2012). This dynamical systems approach has been fruitfully utilized in the Earth climate literature for various climate variables and data-sets (De Luca, et al., 2020a; Hochman , et al., 2021; Rodrigues , et al., 2018; Faranda, et al., 2019a; Faranda, et al., 2019b; De Luca, et al., 2020b). This perspective permits the computation of instantaneous characteristics of chaotic dynamical systems. Hence, it is suitable to investigate the time-series dynamics of exoplanet atmospheres. The temporal sequence of two-dimensional atmospheric variables, in our case *Tmax (*see main text) *or Tmin (see Appendix)*, are used as samples from a long trajectory in the atmosphere's phase-space. For each daily longitude-latitude map, we compute instantaneous dynamical properties. The analysis concentrates on two metrics: persistence ($\theta^{-1}$) and local dimension ($d$) (Faranda, et al., 2017). The persistence ($\theta^{-1}$) of a specific atmospheric state would approximate for how long the *Tmax* maps in the ERA5 reanalysis or ExoCAM model simulations resemble the chosen atmospheric state every time the trajectory arrives near that state. The local dimension ($d$) is an estimate for the number of options that the atmospheric state can transition from and to (see Fig. 2 for some intuition on the dynamical systems metrics meaning).

In practical terms, we consider an atmospheric variable *x* (i.e. *Tmax or Tmin*) over a given domain (e.g., one of the regions in Fig. 1 a) and a state of interest $\zeta$. Then, we compute a value of $\theta^{-1}$ and *d* for each time step of *x* (i.e., each day). Recurrences of $\zeta$ are states that are close to $\zeta$ in phase-space, and thus have spatial configurations that are very similar to $\zeta$ in physical space. We define recurrences using the Euclidean distance (*dist*). To compute recurrences, one has to first define an observable via logarithmic returns as follows:

$$g(x(t), \zeta) = -\log[dist(x(t), \zeta)]$$

Where *x(t)* represents the complete time-series of the variable *x*. Then, a sufficiently high quantile threshold s(q,$\zeta$)(in our case the 98[th] percentile) is defined for the time-series g(x(t),$\zeta$), so that for g(x(t),$\zeta$) > s(q,$\zeta$) (i.e. a recurrence) it is possible to define u($\zeta$)= g(x(t),$\zeta$)- s(q,$\zeta$). The cumulative distribution function F(u,$\zeta$) converges to the exponential member of the Generalized Pareto Distribution, which is relevant for modeling the tails of physical distributions (Freitas, et al., 2010):

$$F(u, \zeta) \simeq exp\left[-\vartheta(\zeta)\frac{u(\zeta)}{\sigma(\zeta)}\right]$$

Where the parameters $\vartheta$ and $\sigma$ are functions of the chosen state $\zeta$. The local dimension (*d*) is then computed as *d = 1/$\sigma$*. The parameter $\vartheta$ is the extremal index and is here estimated using the approach in (Süveges, 2007). While the persistence is given by:

$$\theta^{-1}(\zeta) = \frac{\Delta t}{\vartheta(\zeta)}$$

Where *$\Delta t$* is the time interval between *Tmax or Tmin* maps. In practice, we obtain a value of $\theta^{-1}$ and *d* for each output time step in our dataset. From this point on we use the inverse persistence metric (*$\theta$*) for illustration purpose only, so that when both *$\theta$* and *d* are low there is less change with time of the atmosphere, and vice versa.

## 3.4 STATISTICAL INFERENCE

The statistical significance for the median spatial differences of climate extremes has been assessed for each grid-box with a two-tailed Wilcoxon rank sum test (Mann & Whitney, 1947). In addition to the Wilcoxon rank sum test, we also implemented the Bonferroni correction to the p-values obtained (Bonferroni, 1936; Sedgwick, 2014). The Bonferroni correction considers Type I errors (or false positives) that can occur when performing a large number of statistical tests. Lastly, we tested the significance at the 5% level (p-value < 0.05) of the differences between medians and standard deviations for extreme anomalies and dynamical systems properties. For the medians, we perform a two-tailed Wilcoxon rank sum test for Earth and TRAPPIST-1e between 'Low' and 'Mid', and 'Low' and 'High' $pCO_2$. The same applies when testing the standard deviations, but here we used a bootstrap test with 10,000 realizations (Markowski & Markovski, 1990). The statistical tests we used throughout this study have been earlier used in many Earth climate studies (e.g., De Luca, et al., 2020b; De Luca, et al., 2020a; Hochman, et al., 2022; Hochman, et al., 2021).

## 4.RESULTS

### 4.1   SPATIAL DIFFERENCES BETWEEN TRAPPIST-1e AND EARTH CLIMATE EXTREMES

We first analyze the spatial differences between TRAPPIST-1e and Earth climate extremes. To do so, we start by comparing the TRAPPIST-1e mean climatology to its extremes (Fig. 1 with respect to Fig. 3 a, c, e and Fig. 4 a, c, e, respectively). For *Tmax*, the highest mean temperature (~20°C - 30°C in the 'Low' and 'Mid' $pCO_2$ scenarios, and ~60°C in the 'High' scenario) is located around the sub-stellar region at the planet's dayside. The minimum values (~-60°C in the 'Low', ~-35°C in the 'Mid' and ~45°C in the 'High' scenarios) are found at mid-latitude anti-stellar regions (Fig. 1 a, c, e). The peak extremes for *Tmax* are located at the equatorial sub-stellar region (~-3°C in the 'Low', ~2°C in the 'Mid' and ~35°C in the 'High' $pCO_2$ scenarios; Fig. 3 a, c, e). For precipitation, the highest mean (~12 mm $d^{-1}$ in the 'Low' and 'Mid' $pCO_2$ scenarios, and ~10mm $d^{-1}$ in the 'High' scenario) and extreme (~30 mm $d^{-1}$ in the 'Low', ~45 mm $d^{-1}$ in the 'Mid' and ~55 mm $d^{-1}$ in the 'High' scenarios) values are located at the sub-stellar equatorial region (Fig. 1 b, d, f and Fig. 4 a, c, e). Note however that for the 'High' $pCO_2$ scenario another maximum emerges at the Polar Regions (Figs. 1 f and 4 e). Indeed, the higher the $pCO_2$ and consequently the higher temperatures allow precipitation to develop at the Poles.

Next, we compare the climate extremes of the Earth-analogue simulations to ERA5 reanalysis (*Tmax* in Fig. 3 b, d, f with Fig. 3 g and precipitation in Fig. 4 b, d, f with Fig. 4 g). Both the Earth-analogue and ERA5 show the maxima for extremes of *Tmax* and precipitation at equatorial regions. However, the Earth analogue shows higher absolute values, particularly for *Tmax* (~60°C in the 'Mid' pCO2 scenario and ~70°C in the 'High' scenario compared to ~30°C in ERA5). This is due to the higher $pCO_2$ we consider, as compared to the observed values in present-day Earth (see Sect. 3.1). In addition, the spatial variability in ERA5 is larger than the Earth-analogue, in which *Tmax* ranges from -50°C – 55°C in ERA5 as compared to just ~25°C - 45°C in the Earth-analogue (cf. the 'Low' $pCO_2$ scenario in Fig. 3 b to Fig. 3 g). The reason is that in the Earth-analogue simulation, we consider an aqua-planet with zero obliquity and hence no seasonality (see Sect. 3.1), whereas this is not the case in the representation of Earth in ERA5. Indeed, when considering *Tmax* in ERA5, 'hot spots' are located over land, e.g., over Australia and the mid-latitude desert strip, whereas, for the Earth-analogue the 'hot spot' is uniformly distributed around the equator (Fig. 3, b, d, f and g). The highest precipitation values in ERA5 are situated over the oceans particularly close to the continents and at equatorial regions; however, in the Earth-analogue simulations a very clear Inter-Tropical Convergence Zone (ITCZ) and precipitation patterns resembling the Hadley, mid-latitude and Polar cells are revealed (Fig. 4 b, d, f and g). Note that for the 'Mid' $pCO_2$ scenario a double ITCZ emerges, which is not apparent in the other two scenarios. Such a feature may be linked to differences in the atmospheric energy transport in the 'Mid' scenario (e.g., Adam, et al., 2018; cf. Fig. 4 d to Fig. 4 b and f).

Finally, we quantitatively compare the differences in climate extremes between TRAPPIST-1e, the Earth-analogue simulation and ERA5 (Figs. 5 and 6). In these Figures, Red (Blue) refers to warmer/drier (colder/wetter) extreme conditions on Earth relative to TRAPPIST-1e. Generally, both the Earth-analogue and ERA5 display warmer extremes as compared to TRAPPIST-1e except for the 'High' scenario, in which TRAPPIST-1e displays warmer extremes (Fig. 5 a, b, d, e, g, h). Indeed, the largest differences in $Tmax$ are situated at the mid-latitude anti-stellar regions for the 'Low' and 'Mid' scenarios (Fig. 5 a, b, d, e; ~45°C - 75°C). This is the coldest region in the TRAPPIST-1e simulation due to cold-core Rossby gyres formed because of the planetary Matsuno-Gill wave pattern induced by the contrast in irradiation from dayside to nightside (Pierrehumbert & Hammond, 2019). However, in the 'High' scenario, particularly for the difference between ERA5 and TRAPPIST-1e, the largest differences are located at the southern pole (Fig. 5 h). Regarding precipitation, the Earth-analogue is significantly wetter at equatorial regions, but rather drier at relatively small regions north and south of the sub-stellar point, especially in the 'Low' and 'High' scenarios (Fig. 6 a, g). In the 'High' $pCO_2$ scenario, the Earth-analogue is also drier at Polar Regions (Fig. 6 h).

Shifting the focus to the difference between ERA5 and TRAPPIST-1e climate extremes, the general spatial difference patterns are kept, but the location, sign and numerical values vary (cf. Fig. 6 b, e, h with Fig. 6 a, d, g). As also mentioned above, we relate this to TRAPPIST-1e assumed to be a tidally locked aqua-planet with higher $pCO_2$ and slower rotation rate as observed on Earth (see Sect. 2). Indeed, ERA5 has significantly larger $Tmax$ extremes over the continents, whereas significantly lower $Tmax$ extremes at sub-stellar regions as compared to TRAPPIST-1e (Fig. 5 b, e, h). ERA5 displays wetter (drier) extremes at equatorial regions and at mid-latitude coastal regions (mid-latitude sub-stellar and Polar regions; Fig. 6 b, e, h). As a reference, we further provide the differences in climate extremes between ERA5 and the Earth analogue (Fig. 5 c, f, i and Fig. 6 c, f, i).

## 4.2 SENSITIVITY OF CLIMATE EXTREMES TO CHANGES IN $CO_2$ PARTIAL PRESSURE

The climate sensitivity of TRAPPIST-1e and the Earth-analogue to $pCO_2$ show some significant differences (Tables 1 - 2 and Figs. 7 – 10). In these Figures, red (blue) color refers to TRAPPIST-1e (Earth) and black/grey colors are the ERA5 reanalysis values. Starting from $Tmax$, both TRAPPIST-1e and the Earth-analogue show a significant increase in both the extremes and median values related to an increase in $pCO_2$ (Figs. 7 and 8, Tables 1 and 2).

Globally, the increase in *Tmax* extremes and medians for TRAPPIST-1e is ~1.5 times larger than the increase for the Earth-analogue; indeed, the Earth case exhibits a greater increase in values from 'Low' to 'Mid' $pCO_2$ compared to TRAPPIST-1e. Yet, TRAPPIST-1e exhibits a greater increase in values from 'Mid' to 'High' $pCO_2$ (Figs. 7 a and 8 a). However, observing in more detail we find that the larger sensitivity is particularly located at anti-stellar and terminator regions rather than at sub-stellar ones (Fig. 7 b – g and Fig. 8 b – g). The differences between the regions are most probably related to the fact that the more sensitive regions in TRAPPIST-1e are areas transitioning between temperatures that are well below zero degrees to above zero degrees as a function of the increase in $pCO_2$. Moreover, these regions do not receive irradiation, thus the day to night temperature contrast is decreasing with increasing $pCO_2$.

Next, we describe the influence changes in $pCO_2$ have on precipitation in both planets (Tables 1 and 2; Figs. 9 and 10). Globally, there are some differences between TRAPPIST-1e and the Earth-analogue. Indeed, the TRAPPIST-1e simulations provide evidence for an increase in both the extremes, the median and variability with an increase in $pCO_2$ (Figs. 9 a and 10 a, Tables 1 and 2). However, the Earth-analogue displays a somewhat different picture. The extremes do show a significant increase with $pCO_2$ increase (Fig. 9 a; Table 1), but the median values show a general decrease and an increase in variability (Fig. 10 a; Table 2). This paradoxical relation is also projected for some regions on Earth, i.e., an overall decrease in precipitation, but with an increase in the extremes (IPCC, 2021).

The detailed regional analysis provides further insights into precipitation variability as a function of an increase in $pCO_2$. The increase in extremes for TRAPPIST-1e as $pCO_2$ increases are particularly evident at equatorial sub-stellar and east-terminator regions (Fig. 9 e, g and Table 1). The asymmetry between east and west terminator regions is due to the super-rotating eastward equatorial jet that preferentially transports water vapor lofted upward at the sub-stellar regions to the eastern terminator. The Earth analogue shows the largest variations in extremes at all equatorial regions, i.e., ITCZ regions (Fig. 9 c – f and Table 1). Indeed, the extremes are particularly influenced where the highest amounts of precipitation are located in each planet (Fig. 4).

Figure 10 shows that the median precipitation values of TRAPPIST-1e significantly decrease and variability increases at mid-latitude sub-stellar, equatorial sub-stellar and equatorial anti-stellar regions due to an increase in $pCO_2$ (Fig. 10 c, d, e and Table 2). Indeed, in these regions TRAPPIST-1e displays a ~2 times stronger sensitivity of precipitation to changes in $pCO_2$. However, the Earth-analogue shows significant decreases in median precipitation and an increase in variability in all regions (Fig. 10 b – g and Table 2). The time series for both the annual climate means and extremes are displayed in Figure 11. Generally, the time-series variability in both mean and extreme precipitation is larger than in temperature and increases with increasing mean precipitation.

**4.3 SENSITIVITY OF CLIMATE DYNAMICS TO CHANGES IN CO$_2$ PARTIAL PRESSURE**

Changes in the climate dynamics of TRAPPIST-1e and the Earth-analogue due to an increase in pCO$_2$ are examined using a dynamical systems point of view computed for the *Tmax* variable (Fig. 12 and Table 3). The first evident result is that an increase in pCO$_2$ increases the persistence of the atmospheric circulation (lowering of $\theta$). This effect is much larger in the TRAPPIST-1e simulations as compared to the Earth-analogue simulations, except for the equatorial sub-stellar region. Indeed, in that region of highest solar incoming radiation the persistence is much higher (low $\theta$) as compared to other regions on TRAPPIST-1e (see Fig. 12 left column and Table 3).

The second key finding is that for the Earth-analogue simulations we see a uniform increase in persistence (lowering of $\theta$) both globally and in all regions (Fig. 12 middle column and Table 3). This result has important implications as to the changes in the climate dynamics on Earth due to climate change. Indeed, we show here that an increase in pCO$_2$ may result in an increase in persistent atmospheric configurations including their extremes and therefore a direct change in their possible impacts. In addition, some regions both in TRAPPIST-1e and in the Earth-analogue show a decrease in $d$ with increasing pCO$_2$. Taking the changes in both $\theta$ and $d$ together suggests a tendency towards an atmosphere that changes less with time. This is particularly prominent for equatorial regions in the Earth-analogue and for all regions in TRAPPIST-1e except for the equatorial sub-stellar region (Fig. 12 and Table 3).

Finally, when comparing the Earth-analogue simulations (Fig. 12 middle column and Table 3) to ERA5 reanalysis (Fig. 12 right column and Table 3) we find that the time-series dynamics in the 'real' Earth tends to change more with time (higher $\theta$ and $d$) as compared to the Earth-analogue (note the different ranges of the y-axis in Fig. 12). We relate this to the fact that in the Earth simulation, we consider an aqua-planet with zero obliquity and thus no seasonality (see Sect. 2), whereas this is not the case as Earth is represented in ERA5.

**5. SUMMARY AND CONCLUSIONS**

In this study, we perform a suite of ExoCAM model simulations with varying CO$_2$ partial pressure (pCO$_2$) to make inferences about the sensitivity of TRAPPIST-1e climate extremes, variability and dynamics as compared to an Earth analogue and present-day Earth. The key findings and conclusions are as follows:

1. **We provide evidence of significant spatial differences in climate extremes between TRAPPIST-1e, the Earth analogue and present-day Earth.** These are associated with the elementary planetary parameters such as tidal locking, rotation rate, incident stellar flux, seasonality and location of land and oceans.

2. **The climate of TRAPPIST-1e is more sensitive to changes in pCO$_2$ as compared to the Earth analogue.** Indeed, both the climatology and extremes of *Tmax* display a larger increase for TRAPPIST-1e, depending on the $pCO_2$ regime. When considering precipitation, TRAPPIST-1e is shown to be more sensitive to an increase in $pCO_2$, particularly at locations were most of the precipitation is concentrated, i.e., equatorial sub-stellar and east-terminator regions. Moreover, our Earth simulations present a paradoxical increase in precipitation extremes and decrease in the mean values, which TRAPPIST-1e does not. This finding may be strongly related to key differences in the basic planetary properties (rotation rate and tidal locking) of the two planets. We therefore conclude that the atmosphere of TRAPPIST-1e may be more sensitive to variations in $pCO_2$ than Earth-like planets. This is likely due to the different climatic feedbacks at work on slowly rotating tidally locked rocky planets, for instance the sub-stellar cloud feedback (Yang, et al., 2013). Future work studying a range of tidally locked rocky exoplanets is required to further elucidate the mechanisms that regulate extreme climate behavior in exoplanet atmospheres relative to Earth.

3. **We use a novel approach grounded in dynamical systems theory to test the sensitivity of TRAPPIST-1e climate dynamics as compared to an Earth analogue and present-day Earth.** This alternative point of view suggests that an increase in $pCO_2$ results in an atmosphere that changes less with time, i.e., higher persistence (lower $\theta$) and lower local dimension ($d$). Such a relation has been earlier demonstrated in different regions on Earth using model simulations (Pfleiderer, et al., 2019; Faranda, et al., 2019c), but here we find that it also applies on a global scale. Particularly evident is the higher persistence associated with higher $pCO_2$, which may also result in longer lasting extremes. This perspective provides additional evidence that the dynamical characteristics of TRAPPIST-1e are also more sensitive to $pCO_2$ variations as compared to Earth-like terrestrial bodies.

4. **Our dynamical systems viewpoint may also be an important tool for identifying habitable exoplanets.** Indeed, a key and often overlooked factor of habitability is climate variability (Popp & Eggl, 2017). Notably, plant growth requires inter-annual, seasonal and diurnal cycles to occur. The dynamical systems approach we implement here can provide quantitative and qualitative evidence for such variability. Certainly, when comparing TRAPPIST-1e with the Earth-analogue and ERA5 reanalysis we find significant differences, with the tidally locked TRAPPIST-1e showing enhanced sensitivity of extremes to $pCO_2$. Notably, tidally locked planets orbiting late-type M dwarf stars may have climate variability at the level that is usually produced by seasons, even though the planet has zero obliquity. However, a full analysis of this issue is out of the scope of the present study and we plan to pursue this avenue in the near future.

With the advent of the James Webb Space Telescope (JWST) and extremely large ground-based observatories, there is a promise that the atmospheres of rocky exoplanets shall be characterized in detail for the first time. Indeed, JWST may be able to constrain $pCO_2$ for TRAPPIST-1e (May, et al., 2021), which would provide a rough understanding of the expected climatic state. However, deciding which exoplanets ought to be studied is *a priori* unknown (Lingam & Loeb, 2018). Our pathway for simulating the climate sensitivity to different planetary parameters including atmospheric composition can help determine which planets will have 'stable' climates that are conducive to life.

As a caveat, we note that we assume that TRAPPIST-1e is tidally locked, whereas it may be in a higher-order spin-orbit resonance (Leconte, et al., 2015; Turbet, et al., 2018). Moreover, our interpretation depends on one model and one simulation per $pCO_2$ scenario. To fully characterize the uncertainty of exoplanet climate extremes, dynamics and habitability, a large ensemble of model simulations varying additional planetary parameters is foreseen.

The novel perspective presented here, which combines a dynamical systems approach with traditional techniques to investigate climate extremes, dynamics and habitability of exoplanets, provides insight into how the diversity of exoplanet properties affects their climates, and it can be directly applied to other terrestrial bodies in our solar system and beyond.

**ACKNOWLEDGEMENTS** A.H thanks the German Helmholtz Association 'Changing Earth' program for funding, and the Hebrew University of Jerusalem for providing technical support. The model simulations were completed with resources provided by the University of Chicago Research Computing Centre (PI: Jacob Bean). We would like to thank the editor and an anonymous Reviewer for providing helpful comments and suggestions, which definitely helped to improve our manuscript.

**DATA AVAILABILITY STATEMENT** The analysis in this paper is based on the European Centre for Medium-range Weather Forecasting (ECMWF) ERA5 reanalysis (https://www.ecmwf.int/en/forecasts/datasets/reanalysis-datasets/era5; Hersbach et al., 2020) and ExoCAM model simulations available upon request from TDK. The code we have used for our dynamical systems analysis is freely available at this location: https://es.mathworks.com/matlabcentral/fileexchange/95768-attractor-local-dimension-and-local-persistence-computation.

**REFERENCES**

Adam, O., Schneider, T. & Brient, F., 2018. Regional and seasonal variations of the double-ITCZ bias in CMIP5 models. *Climate Dynamics,* Volume 51, pp. 101-117.

Ajjur, S. & Al-Ghamdi, S., 2021. Global hotspots for future absolute temperature extremes from CMIP6 models. *Earth and Space Science,* Volume 8, p. e2021EA001817.

Allard, F. et al., 2007. K-H2 quasi-molecular absorption detected in the T-dwarf Indi Ba. *Astronomy and Astrophysics,* 474(2), pp. L21-L24.

Bitz, C. et al., 2012. Climate sensitivity of the community climate system model, version 4. *Journal of Climate,* 25(9), pp. 3053-3070.

Bonferroni, C., 1936. Teoria statistica dell classi e calcolo delle probabilita. *Publicazioni Del R Istituto Superiore Di Scienze Economiche Commerciali Di Firenze,* Volume 8, pp. 3-62.

Charnay, B. et al., 2021. Formation and dynamics of water clouds on temperate sub-Neptunes: the example of K2-18b. *Astronomy and Astrophysics,* Volume 646, p. A171.

Colose, C., Del Genio , A. & Way , M., 2019. Enhanced habitability on high obliquity bodies near the outer edge of the habitable zone of Sun-like stars. *The Astrophysical Journal,* 884(2), p. 138.

De Luca, P. et al., 2020b. Compound warm-dry and cold-wet events over the Mediterranean. *Earth System Dynamics,* Volume 11, pp. 793-805.

De Luca, P., Messori, G., Pons, F. & Faranda, D., 2020a. Dynamical systems theory sheds new light on compound climate extremes in Europe and Eastern North America. *Quarterly Journal of the Royal Meteorological Society,* Volume 146, pp. 1636-1650.

de vries, H., Haarsma, R. & Hazeleger, W., 2012. Western European cold spells in current and future climate. *Geophysical Research Letters,* Volume 39, p. L04706.

Del Genio, A. et al., 2019. Climates of warm Earth-like planets. III. Fractional habitability from a water cycle perspective. *The Astrophysical Journal,* 887(2), p. 197.

Dobbs-Dixon, I., Cumming, A. & Lin , D., 2010. Radiative hydrodynamic simulations of HD209458b: temporal variability. *The Astrophysical Journal,* 710(2), p. 1395.

Easterling, D., Kunkel, K., Wehner, M. & Sun, L., 2016. Detection and attribution of climate extremes in the observed record. *Weather and Climate Extremes,* Volume 11, pp. 17-27.

Emanuel, K., 1988. The minimum intensity of Hurricanes. *Journal of the Atmospheric Sciences,* 45(7), pp. 1143-1155.

Eyring, V. et al., 2016. Overview of the Coupled Model Intercomparison Project Phase 6 (CMIP6) experimental design and organization. *Geoscientific Model Development,* 9(5), pp. 1937-1958.

Faranda, D. et al., 2019a. The hammam effect or how a warm ocean enhances large scale atmospheric predictability. *Nature Communications,* Volume 10, p. 1316.

Faranda, D. et al., 2019c. The hammam effect or how a warm ocean enhances large scale atmospheric predictability. *Nature Communications,* Volume 10, p. 1316.

Faranda, D., Messori, G. & Vannistem, S., 2019b. Attractor dimension of time-averaged climate observables: insights from a low-order ocean-atmosphere model. *Tellus A,* 71(1), p. 1554413.

Faranda, D., Messori, G. & Yiou, P., 2017. Dynamical proxies of North Atlantic predictability and extremes. *Scientific Reports,* Volume 7, p. 41278.

Fauchez, T. et al., 2021. TRAPPIST habitable atmosphere intercomparison (THAI) workshop report. *The Planetary Science Journal,* 2(3), p. 106.

Fauchez, T. et al., 2019. Impact of clouds and hazes on the simulated JWST transmission spectra of habitable zone planets in the TRAPPIST-1 System. *The Astrophysical Journal,* 887(2), p. 194.

Freitas, A., Freitas, J. & Todd, M., 2010. Hitting time statistics and extreme value theory. *Probability theory and related fields,* Volume 147, pp. 675-710.

Giorgi, F., Jones, C. & Asrar, G., 2009. ddressing climate information needs at the regional level: the CORDEX framework. *World Meteorological Organisation Bulletin,* Volume 58, pp. 175-183.

Haqq-Misra, J. et al., 2018. Demarcating circulation regimes of synchronously rotating terrestrial planets within the habitable zone. *The Astrophysical Journal,* 852(2), p. 67.

Haqq-Misra, J. et al., 2019. Constraining the magnitude of climate extremes from time-varying instellation on a circumbinary terrestrial planet. *Journal of Geophysical Research Planets,* Volume 124, pp. 3231-3243.

Hersbach, H. et al., 2020. The ERA5 golabal reanalysis. *Quarterly Journal of the Royal Meteorological Society,* 146(730), pp. 1999-2049.

Hochman , A. et al., 2021. Do Atlantic-European weather regimes physically exist?. *Geophysical Research Letters,* Volume 48, p. e2021GL095574.

Hochman, A. et al., 2019. A new dynamical systems perspective on atmospheric predictability: eastern Mediterranean weather regimes as a case study. *Science Advances,* 5(6), p. eaau0936.

Hochman, A. et al., 2020. The dynamics of cyclones in the twentyfirst century; the eastern mediterranean as an example. *Climate Dynamics,* 54(1), pp. 561-574.

Hochman, A. et al., 2021. A new view of heat wave dynamics and predictability over the eastern Mediterranean. *Earth System Dynamics,* Volume 12, p. 133-149.

Hochman, A. et al., 2022. Dynamics and predictability of cold spells over the eastern Mediterranean. *Climate Dynamics,* Volume 58, pp. 2047-2064.

IPCC, 2021. *The Physical Science Basis. Contribution of Working Group I to the Sixth Assessment Report of the Intergovernmental Panel on Climate Change.* Cambridge: Cambridge University Press.

Jansen, T., Scharf, C., Way, M. & Del Genio, A., 2019. Climates of warm Earth-like planets. II. Rotational "Goldilocks" zones for fractional habitability and silicate weathering. *The Astrophysical Journal,* 875(2), p. 79.

Kaspi, Y. & Showman, A., 2015. Atmospheric dynmaics of terrestrial exoplanets over a wide range of orbital and atmopsheric parameters. *The Astrophysical Journal,* 804(1), p. 60.

Komacek, T. & Abbot, D., 2019. The atmospheric circulation and climate of terrestrial planets orbiting Sun-like and M Dwarf stars over a broad range of planetary parameters. *The Astrophysical Journal,* 871(2), p. 245.

Komacek, T., Chavas, D. & Abbot, D., 2020. Hurricane genesis is favorable on terrestrial exoplanets orbiting late-type M Dwarf stars. *The Astrophysical Journal,* 898(2), p. 115.

Komacek, T. & Showman, A., 2020. Temporal variability in hot Jupiter atmospheres. *The Astrophysical Journal,* 888(1), p. 2.

Kopparapu, R. et al., 2017. Habitable moist atmospheres on terrestrial planets near the inner edge of the habitable zone around M dwarfs. *The Astrophysical Journal,* 845(1), p. 5.

Krissansen-Totton, J., Garland, R., Irwin, P. & Catling, D., 2018. Detectability of biosignatures in anoxic atmospheres with the James Webb Space Telescope: A TRAPPIST-1e case study. *The Astronomical Journal,* 156(3), p. 114.

Leconte, J., Wu, H., Menou, K. & Murray, N., 2015. Asynchronous rotation of Earth-mass planets in the habitable zone of lower-mass stars. *Science,* 347(6222), pp. 632-635.

Lingam, M. & Loeb, A., 2018. Optimal target stars in the search for life. *The Astrophysical Journal Letters,* Volume 857, p. L17.

Lucarini, V., Faranda, D. & Wouters, J., 2012. Universal behavior of extreme value statistics for selected observables of dynamical systems. *Journal of Statistical Physics,* Volume 147, pp. 63-73.

Lustig-Yaeger, J., Meadows, V. & Linkowski, A., 2019. The detectability and characterization of the TRAPPIST-1 exoplanet atmospheres with JWST. *The Astronomical Journal,* 158(1), p. 27.

Mann, H. & Whitney, D., 1947. On a test whether two random variables is stochastically larger than the other. *Annals of Methematical Statistics,* 18(1), pp. 50-60.

Markowski, C. & Markovski, E., 1990. Conditions for the effectiveness of a preliminary test of variance. *The American Statistician,* 44(4), pp. 322-326.

May, E. et al., 2021. Water ice cloud variability and multi-epoch transmission spectra of TRAPPIST-1e. *The Astrophysical Journal Letters,* 911(2), p. L30.

Meehl, G. et al., 2007. The WCRP CMIP3 multimodel dataset: A new era in climate change research. *Bulletin of the American Meteorological Society,* Volume 88, pp. 1383-1394.

Messori, G., Caballero, R. & Faranda, D., 2017. A dynamical systems approach to studying midlatitude weather extremes. *Geophysical Research Letters,* Volume 44, pp. 3346-3354.

Mikal-Evans, T., 2022. Detecting the proposed CH4-CO2 biosignature pair with the James Webb Space Telescope: TRAPPIST-1e and the effect of cloud/haze. *Monthly Notices of the Royal Astronomical Society,* 510(1), pp. 980-991.

Mukherjee, S., Mishra, A. & Trenberth, K., 2018. Climate change and drought: A perspective on drought indices. *Current Climate Change Reports,* Volume 4, pp. 145-163.

Neale, R. et al., 2010. *Description of the NCAR community atmosphere model (CAM 5.0).,* Boulder, Colorado: NCAR Tech.

Noda, S. et al., 2017. The circulation pattern and day-night heat transport in the atmosphere of a synchronously rotating aquaplanet: Dependence on planetary rotation rate. *Icarus,* Volume 282, pp. 1-18.

PAGES2k consortium, 2017. A global multiproxy database for temperature reconstructions of the Common Era. *Scientific Data,* Volume 4, p. 170088.

Pfleiderer, P., Schleussner, C., Kornhuber, K. & Koumou, D., 2019. Summer weather becomes more persistent in a 2°C world. *Nature Climate Change,* Volume 9, pp. 666-671.

Pierrehumbert, R. & Hammond, M., 2019. Atmospheric circulation of tide-locked exoplanets. *Annual Review of Fluid Mechanics,* Volume 51, pp. 275-303.

Popp, M. & Eggl, S., 2017. Climate variations on Earth-like circumbinary planets. *Nature Communications,* Volume 8, p. 14957.

Rauscher , E. et al., 2007. Hot Jupiter variabiliy in eclipse depth. *The Astrophsical Journal,* 662(2), p. L115.

Rodrigues , D. et al., 2018. Dynamical Properties of the North Atlantic Atmospheric Circulation in the Past 150 Years in CMIP5 Models and the 20CRv2c Reanalysis. *Journal of Climate,* 31(15), pp. 6097-6111.

Schewe, J. et al., 2019. State-of-the-art global models underestimate impacts from climate extremes. *Nature Communications,* Volume 10, p. 1005.

Sedgwick, P., 2014. Multiple hypothesis testing and Bonferroni's correction. *British Medical Journal,* Volume 349, p. g6284.

Sergeev, D. et al., 2021. The TRAPPIST-1 habitable atmosphere intercomparison (THAI). Part II: Moist cases - The two waterworlds. *The Planetary Science Journal.*

Sillmann, J. et al., 2013. Climate extremes indices in the CMIP5 multimodel ensemble: Part 2. Future climate projections. *Journal of Geophysical Research Atmospheres,* Volume 118, pp. 2473-2493.

Stott, P., Stone, D. & Allen, M., 2004. Human contribution to the European heatwave of 2003. *Nature,* Volume 432, pp. 610-614.

Suissa, G. et al., 2020. Dim prospects for transmission spectra of ocean Earths around M stars. *The Astrophysical Journal,* 891(1), p. 58.

Süveges, M., 2007. Likelihood estimation of the extremal index. *Extremes,* 10(1-2), p. 41-55.

Taylor, K., Stouffer, G. & Meehl, G., 2012. An overview of CMIP5 and the experiment design. *Bulletin of the American Meteorological Society,* Volume 93, pp. 485-498.

Turbet, M. et al., 2018. Modeling climate diversity, tidal dynamics and the fate of volatiles on TRAPPIST-1 planets. *Astronomy and Astrophysics,* Volume 612, p. A86.

Turbet, M. et al., 2021. The TRAPPIST-1 habitable atmosphere intercomparison (THAI). Part I: Dry cases -- The fellowship of the GCMs. *the Planetary Science Journal.*

Vogel, E. et al., 2019. The effects of climate extremes on global agricultural yields. *Environemntal Research Letters,* Volume 14, p. 054010.

Way, M. et al., 2018. Climates of warm Earth-like planets. I. 3D model simulations. *The Astrophysical Journal,* 239(2), p. 24.

Wolf, E., 2017. Assessing the habitability of the TRAPPIST-1 system using a 3D climate model. *The Astrophysical Journal Letters,* 839(1), p. L1.

Wolf, E., Kopparapu, R., Haq-Misra, J. & Fauchez, T., 2022. ExoCAM: A 3D climate model for exoplanets atmospheres. *Tha Planetary Science Journal,* 3(7), p. 17.

Wolf, M. et al., 2021. Possible substellar companions in dwarf eclipsing binaries. *Astronomy and Astrophysics,* Volume 647, p. A65.

Yang, H., Komacek, T. & Abbot, D., 2019. Effects of radius and gravity on the inner edge of the habitable zone. *The Astrophysical Journal Letters,* 87(2), p. L27.

Yang, J. et al., 2019. Ocean dynamics and the inner edge of the habitable zone for tidally locked terrestrial planets. *The Astrophysical Journal,* 871(1), p. 29.

Yang, J., Cowan, N. & Abbot, D., 2013. Stabilizing cloud feedback dramatically expands the habitable zone of tidally locked planets. *The Astrophysical Journal Letters,* 771(2), p. L45.

Yan, M. & Yang , J., 2020. Hurricanes on tidally locked terrestrial planets: fixed surface temperature experiments. *Astronomy and Astrophysics,* Volume 643, p. A37.

Zelinka, M. et al., 2020. Causes of Higher Climate Sensitivity in CMIP6 Models. *Geophysical Research Letters,* Volume 47, p. e2019GL085782.

**Table 1** Summary of changes between the Earth-analogue and TRAPPIST-1e median of *extreme anomalies* with respect to ERA5 for daily maximum temperature (*Tmax* in °C) and total precipitation (mm d$^{-1}$, Figs. 7 and 9, respectively). The different pCO$_2$ scenarios are described in Section 3.1 and are referred to as 'Low' – 10$^{-2}$ Bar, 'Mid' – 10$^{-1}$ Bar and 'High' – 1Bar. Significant differences according to the Wilcoxon rank sum test for the medians and bootstrap test (n=10,000) for the variances at the 5% level as compared to the 'Low' pCO$_2$ scenario are marked with an Asterix (*). Standard deviations are shown in brackets. The considered regions are exhibited in Fig. 1 a. ERA5 reanalysis values are displayed for reference.

| CO$_2$ | Low | Mid | High | Low | Mid | High |
|---|---|---|---|---|---|---|
| Variable | | Tmax (°C) | | | Precipitation (mm d$^{-1}$) | |
| Region | | | Global | | | |
| ERA5 | 1.96 (1.76) | | | 10.82 (9.99) | | |
| Earth | 16.78 (10.48) | 49.75* (14.85*) | 68.05* (16.37*) | 15.35 (30.78) | 37.31* (21.39*) | 30.98* (19.19*) |
| TRAPPIST-1e | -17.30 (22.14) | -8.41* (21.56*) | 40.80* (19.07*) | 1.22 (5.48) | 3.63* (8.11*) | 22.92* (33.85*) |
| Region | | | Mid-latitude anti-stellar | | | |
| ERA5 | 2.27 (1.92) | | | 12.60 (5.06) | | |
| Earth | 11.20 (2.36) | 45.99* (2.92) | 63.00* (4.15*) | 19.19 (5.67) | 32.66* (1.54*) | 23.84* (2.72*) |
| TRAPPIST-1e | -44.05 (6.37) | -29.14* (7.05) | 30.33* (4.30*) | -0.70 (0.96) | 0.75* (0.86) | 4.82* (10.30*) |
| Region | | | Mid-latitude sub-stellar | | | |
| ERA5 | 1.41 (0.53) | | | 21.13 (5.63) | | |
| Earth | 16.28 (2.32) | 51.58* (3.93*) | 68.29* (4.84*) | 18.52 (5.84) | 32.23* (2.13*) | 22.70* (4.14*) |
| TRAPPIST-1e | -1.00 (1.98) | 7.05* (1.95) | 47.73* (4.63*) | 5.57 (2.81) | 13.72* (3.34) | 8.80* (5.91*) |
| Region | | | Equatorial anti-stellar | | | |
| ERA5 | 1.13 (0.84) | | | 11.18 (8.43) | | |
| Earth | 18.24 (4.38) | 42.43* (3.52*) | 58.47* (3.72*) | 40.21 (59.27) | 69.48* (15.63*) | 63.51* (17.99*) |
| TRAPPIST-1e | -42.19 (5.05) | -32.34* (5.06) | 23.36* (3.80*) | 1.30 (2.17) | 3.50* (2.15) | 7.39* (2.15) |
| Region | | | Equatorial sub-stellar | | | |
| ERA5 | 1.11 (0.18) | | | 29.63 (8.07) | | |
| Earth | 16.21 (1.30) | 41.06* (0.71*) | 57.73* (0.67*) | 36.47 (61.25) | 69.95* (19.93*) | 58.01 (18.75*) |
| TRAPPIST-1e | -3.34 (1.80) | 2.23* (1.45*) | 36.06* (0.85*) | 28.59 (10.53) | 43.95* (11.72) | 52.51* (6.26*) |
| Region | | | Equatorial west-terminator | | | |
| ERA5 | 1.02 (0.49) | | | 24.89 (5.04) | | |
| Earth | 16.32 (2.09) | 41.27* (1.14*) | 57.50* (1.43*) | 37.83 (60.35) | 64.76* (15.77*) | 60.49* (18.22*) |
| TRAPPIST-1e | -45.20 (5.95) | -31.52* (2.48*) | 24.50* (1.57*) | -4.52 (1.74) | -3.93 (1.93) | -1.50* (2.85*) |
| Region | | | Equatorial east-terminator | | | |
| ERA5 | 1.31 (0.87) | | | 15.09 (18.04) | | |
| Earth | 18.44 (3.13) | 43.41* (2.61) | 59.25* (2.92) | 38.37 (60.75) | 79.53* (20.00*) | 61.79* (19.18*) |
| TRAPPIST-1e | -31.09 (3.89) | -25.62* (3.35) | 24.87* (2.99*) | -1.59 (3.43) | -0.84* (3.60) | 31.20* (14.25*) |

**Table 2** Summary of changes in the Earth and TRAPPIST-1e *median* maximum temperature (*Tmax* in °C) and precipitation (mm d$^{-1}$) due to an increase in pCO$_2$ (Figs. 8 and 10, respectively). The different pCO$_2$ scenarios are described in Section 3.1 and are referred to as 'Low' – 10$^{-2}$ Bar, 'Mid' – 10$^{-1}$ Bar and 'High' – 1 Bar. Significant differences according to the Wilcoxon rank sum test for the medians and bootstrap test (n=10,000) for the variances at the 5% level as compared to the 'Low' pCO$_2$ scenario are marked with an Asterix (*). Standard deviations are shown in brackets. The considered regions are exhibited in Fig. 1 a. ERA5 reanalysis values are displayed for reference.

| CO$_2$ | Low | Mid | High | Low | Mid | High |
|---|---|---|---|---|---|---|
| Variable | Tmax (°C) | | | Precipitation (mm d$^{-1}$) | | |
| Region | Global | | | | | |
| ERA5 | 16.26 (1.43) | | | 2.92 (0.13) | | |
| Earth | 30.96 (0.26) | 61.12*(0.44*) | 79.49*(0.6*) | 4.57(0.31) | 4.69*(1.08*) | 3.6*(1.21*) |
| TRAPPIST-1e | -24.6(3.31) | -9.81*(2.11*) | 49.57*(0.4*) | 1.00(0.29) | 1.28*(0.41*) | 1.37*(0.97*) |
| Region | Mid-latitude anti-stellar | | | | | |
| ERA5 | 16.13 (5.06) | | | 1.81 (1.07) | | |
| Earth | 26.77 (0.42) | 60.03*(0.6*) | 78.02* (0.64*) | 2.55 (2.39) | 2.87*(3.93*) | 1.8*(2.61*) |
| TRAPPIST-1e | -58.4 (12.79) | -36.92*(11.45*) | 44.48*(0.82*) | 0.04 (0.24) | 0.12*(0.47*) | 0.09*(2.32*) |
| Region | Mid-latitude sub-stellar | | | | | |
| ERA5 | 10.82 (3.61) | | | 3.65 (1.77) | | |
| Earth | 26.82 (0.43) | 60*(0.61*) | 78.13*(0.64*) | 2.53 (2.38) | 2.94*(4.03*) | 1.82*(2.49*) |
| TRAPPIST-1e | 8.86 (1.08) | 18.16*(0.77*) | 55.96*(0.71*) | 1.63 (1.66) | 2.39*(2.95*) | 0.09*(3.83*) |
| Region | Equatorial anti-stellar | | | | | |
| ERA5 | 27.54 (1.73) | | | 2.25 (1.05) | | |
| Earth | 42.37 (0.5) | 66*(1.07*) | 83.3*(1.39*) | 6.76(8.62) | 2.64*(12.33*) | 2.42*(13.91*) |
| TRAPPIST-1e | -31.31(6.06) | -17.45*(4.52*) | 47.33*(0.85*) | 0.45 (0.59) | 0.36*(0.96*) | 0.08*(2.35*) |
| Region | Equatorial sub-stellar | | | | | |
| ERA5 | 27.7 (0.37) | | | 5.95 (2.11) | | |
| Earth | 42.4 (0.41) | 65.86*(1*) | 83.3*(1.41*) | 6.91(8.73) | 2.54*(13.36*) | 2.38*(13.36*) |
| TRAPPIST-1e | 21.43 (0.82) | 26.67*(1.01*) | 60.28*(1.18*) | 11.42(6.94) | 12.21*(9.38*) | 0.78*(10.62*) |
| Region | Equatorial west-terminator | | | | | |
| ERA5 | 27.68 (0.59) | | | 5.76 (2.27) | | |
| Earth | 42.32 (0.49) | 65.88*(1.03*) | 83.36*(1.4*) | 6.83 (8.79) | 2.61*(12.5*) | 2.42*(13.83*) |
| TRAPPIST-1e | -27.76 (5.4) | -10.49*(4*) | 49.93*(1.01*) | 0.05 (0.29) | 0.03*(0.38*) | 0.05*(1.2*) |
| Region | Equatorial east-terminator | | | | | |
| ERA5 | 24.65 (1.17) | | | 3.84 (1.57) | | |
| Earth | 42.43 (0.41) | 65.89*(1.02*) | 83.29*(1.41*) | 6.85 (8.69) | 2.66*(13.94*) | 2.37*(13.79*) |
| TRAPPIST-1e | -16.27 (4.21) | -5.4*(2.3*) | 47.79*(0.63*) | 0.09 (0.21) | 0.04*(0.36*) | 0.11*(6.05*) |

**Table 3** Summary of changes in the Earth and TRAPPIST-1e median *d* (local dimension) and *θ* (inverse persistence) due to an increase in $pCO_2$ (Fig. 12). The different $pCO_2$ scenarios are described in Section 3.1 and are referred to as 'Low' – $10^{-2}$ Bar, 'Mid' – $10^{-1}$ Bar and 'High' – 1 Bar. Significant differences according to the Wilcoxon rank sum test for the medians and bootstrap test (n=10,000) for the variances at the 5% level as compared to the 'Low' $pCO_2$ scenario are marked with an Asterix (*). Standard deviations are shown in brackets. The considered regions are exhibited in Fig. 1 a. ERA5 reanalysis values are displayed for reference.

| $CO_2$ | Low | Mid | High | Low | Mid | High |
|---|---|---|---|---|---|---|
| **Variable** | | **d** | | | **θ** | |
| **Region** | | | **Global** | | | |
| ERA5 | 18.67 (2.66) | | | 0.27(0.05) | | |
| Earth | 4.54(0.99) | 6.51*(1.18*) | 5.31*(1.07*) | 0.05(0.06) | 0.04(0.01) | 0.03(0) |
| TRAPPIST-1e | 8.65(1.44) | 7.53*(1.44) | 3.71*(0.91*) | 0.7(0.1) | 0.78*(0.12) | 0.02*(0) |
| **Region** | | | **Mid-latitude anti-stellar** | | | |
| ERA5 | 8.23(1.16) | | | 0.52(0.07) | | |
| Earth | 5.79(1) | 6.58*(1.17*) | 6.14*(1.06) | 0.12(0.03) | 0.13(0.03) | 0.1(0.02) |
| TRAPPIST-1e | 4.6(0.86) | 4.27*(0.85) | 2.87*(0.89) | 0.81(0.13) | 0.88(0.13) | 0.03*(0.01) |
| **Region** | | | **Mid-latitude sub-stellar** | | | |
| ERA5 | 8.05(1.21) | | | 0.55(0.12) | | |
| Earth | 5.57(0.97) | 6.41*(1.16*) | 6.08*(1.13*) | 0.12(0.03) | 0.13(0.03) | 0.1(0.02) |
| TRAPPIST-1e | 5.66(1.17) | 5.41*(0.93*) | 4.28*(0.91*) | 0.71(0.16) | 0.16*(0.03*) | 0.08*(0.02*) |
| **Region** | | | **Equatorial anti-stellar** | | | |
| ERA5 | 7.93(1.51) | | | 0.47(0.11) | | |
| Earth | 5.91(1.08) | 4.53*(0.86*) | 4.14*(0.98*) | 0.1(0.02) | 0.06(0.01) | 0.05(0.02) |
| TRAPPIST-1e | 7.5(1.41) | 7.59(1.4) | 2.39*(0.86*) | 0.72(0.08) | 0.73(0.08) | 0.02*(0.01*) |
| **Region** | | | **Equatorial sub-stellar** | | | |
| ERA5 | 13.04(1.96) | | | 0.62(0.09) | | |
| Earth | 6(1.01) | 4.48*(0.88*) | 4.03*(0.93*) | 0.1(0.02) | 0.07(0.02) | 0.06*(0.02) |
| TRAPPIST-1e | 6(1.1) | 5.93(1.15) | 5.02*(0.99) | 0.14(0.03) | 0.14(0.03) | 0.08*(0.02) |
| **Region** | | | **Equatorial west-terminator** | | | |
| ERA5 | 9.96(1.52) | | | 0.5(0.08) | | |
| Earth | 5.89(1.1) | 4.4*(0.86*) | 4.03*(0.95) | 0.1(0.02) | 0.07(0.02) | 0.06*(0.02) |
| TRAPPIST-1e | 6.14(1.97) | 6.46*(1.97) | 2.7*(0.93*) | 0.73(0.09) | 0.78(0.11) | 0.03*(0) |
| **Region** | | | **Equatorial east-terminator** | | | |
| ERA5 | 8.83(1.32) | | | 0.44(0.07) | | |
| Earth | 6.1(1.09) | 4.42*(0.82*) | 3.96*(0.96*) | 0.11(0.02) | 0.07(0.01) | 0.06*(0.02) |
| TRAPPIST-1e | 7.2(1.41) | 7.26(1.44) | 2.87*(0.87*) | 0.58(0.08) | 0.57(0.08) | 0.04*(0.01*) |

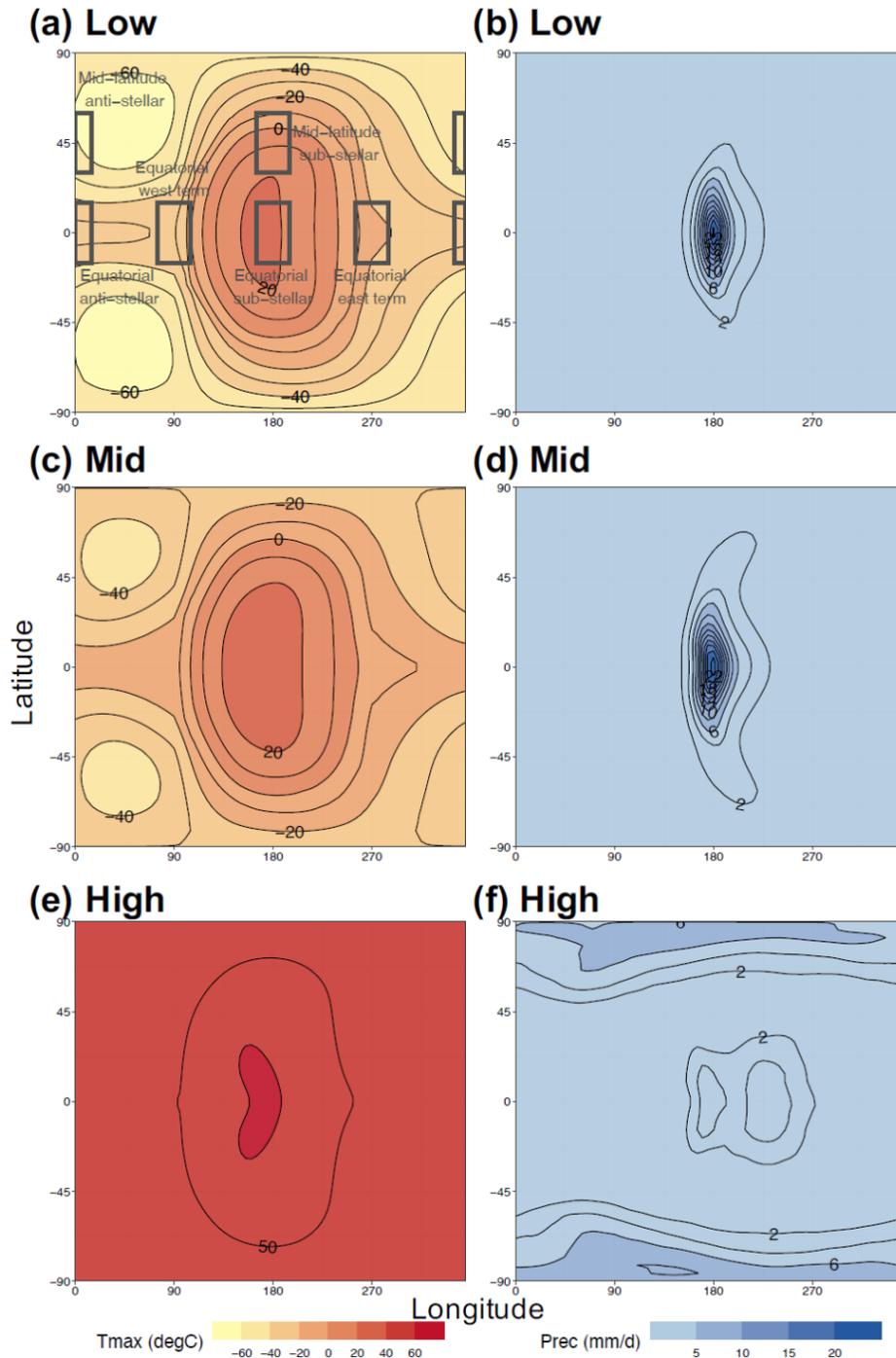

**Figure 1** Climatology of (a, c, e) maximum temperature (*Tmax* in °C) and (b, d, f) total precipitation (*Prec* in mm d$^{-1}$) for TRAPPIST-1e. The climatology is computed using the mean of 80-year ExoCAM simulations with varying pCO$_2$. The different pCO$_2$ scenarios are described in Section 3.1 and are referred to as 'Low' – 10$^{-2}$ Bar, 'Mid' – 10$^{-1}$ Bar and 'High' – 1 Bar. In panel (a) the sub-regions used in Figures 7 – 10 and 12 are marked with grey rectangles. The sub regions considered are global (-90 – 90N, 0 – 360E), mid-latitude anti-stellar (30 – 60N, 345 – 15E), mid-latitude sub-stellar (30 – 60N, 165-195E), equatorial anti-stellar (-15 – 15N, 345 – 15E), equatorial sub-stellar (-15 – 15N, 165 – 195E), equatorial west-terminator (-15 – 15N, 75 – 105E) and equatorial east-terminator (-15 – 15N, 255 – 285E).

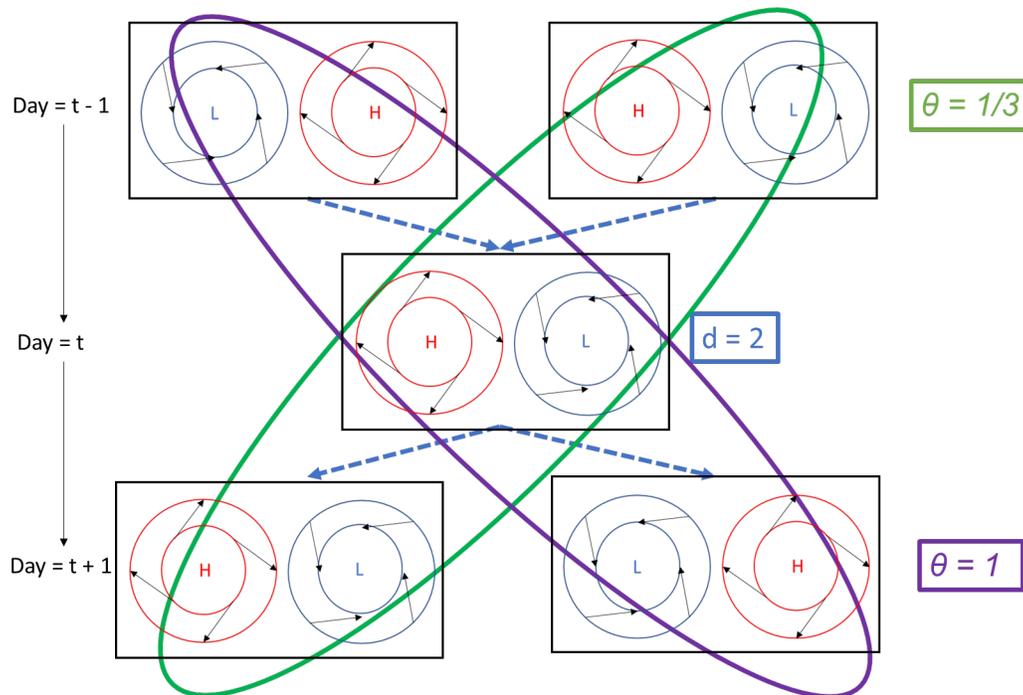

**Figure 2** Schematic representation of the dynamical systems metrics on made-up atmospheric states. The local dimension (*d*) is related to the number of possible atmospheric patterns preceding and following the state being analyzed (in this case *d* = 2), and *θ* is the inverse of the persistence. If the pattern persists for three days (green ellipse), then *θ* = 1/3. If the patterns change at each time step (purple ellipse), then *θ* = 1. Inspired by a figure from (Rodrigues , et al., 2018).

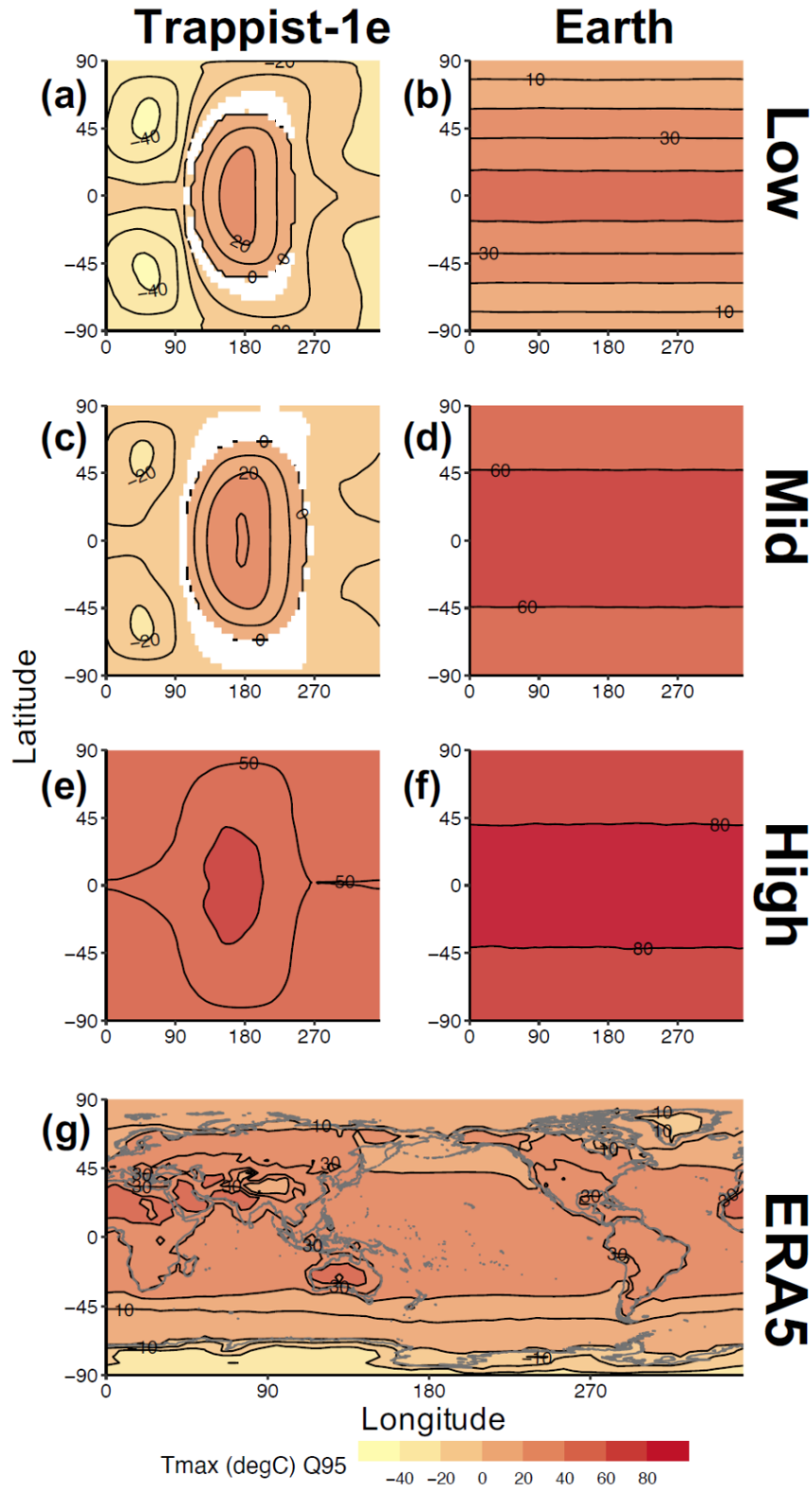

**Figure 3** Means of extreme maximum temperatures (*Tmax* in °C) for (a, c, e) TRAPPIST-1e, (b, d, f) Earth analogue and (g) ERA5 reanalysis. Extremes are daily values exceeding the 95[th] percentile of the whole distribution. The different $pCO_2$ scenarios are described in Section 3.1 and are referred to as 'Low' – $10^{-2}$ Bar, 'Mid' – $10^{-1}$ Bar and 'High' – 1 Bar. Black lines represent isolines of *Tmax* extremes. White regions represent areas without extremes.

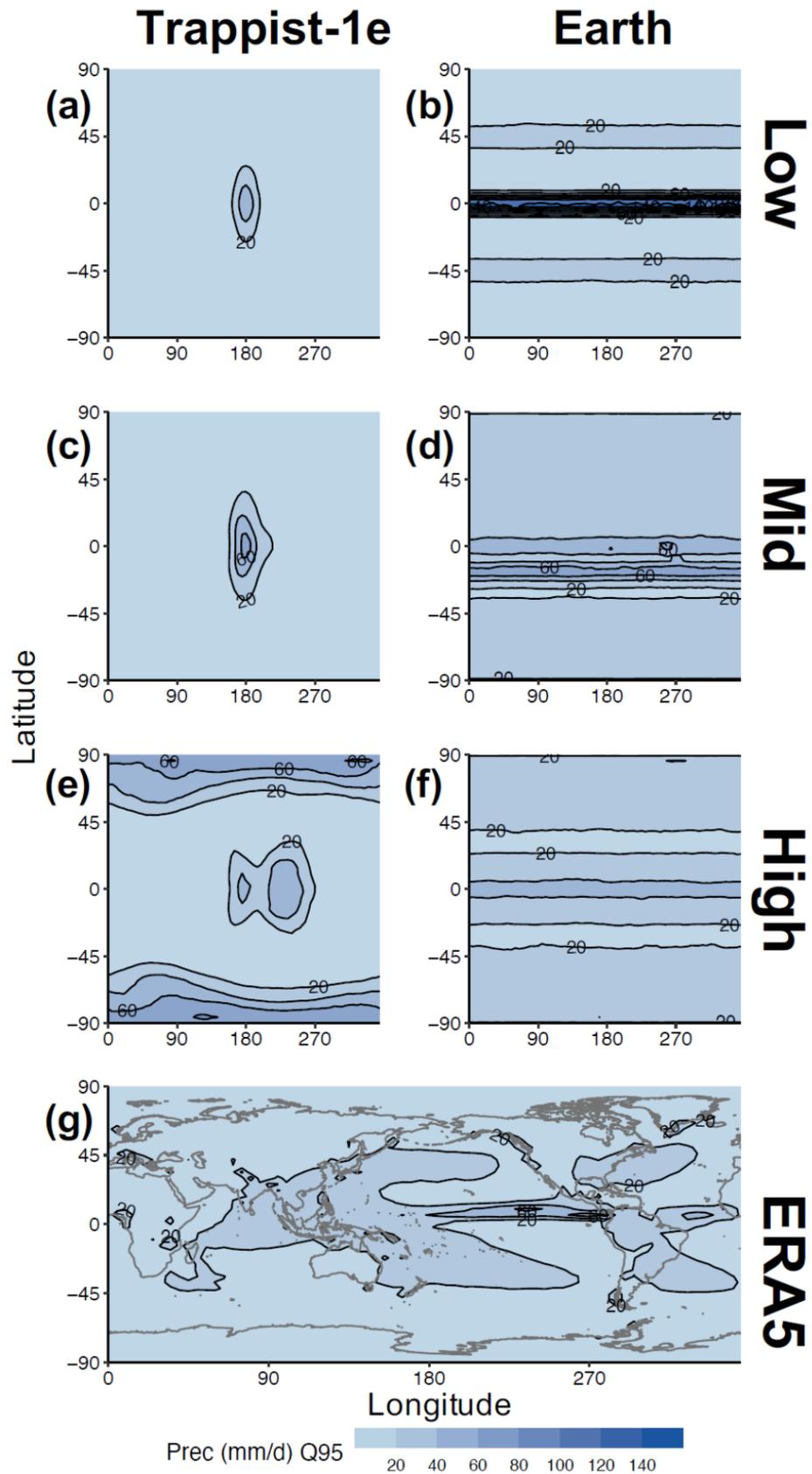

**Figure 4** Same as Fig. 3 but for precipitation (*Prec* in mm d$^{-1}$).

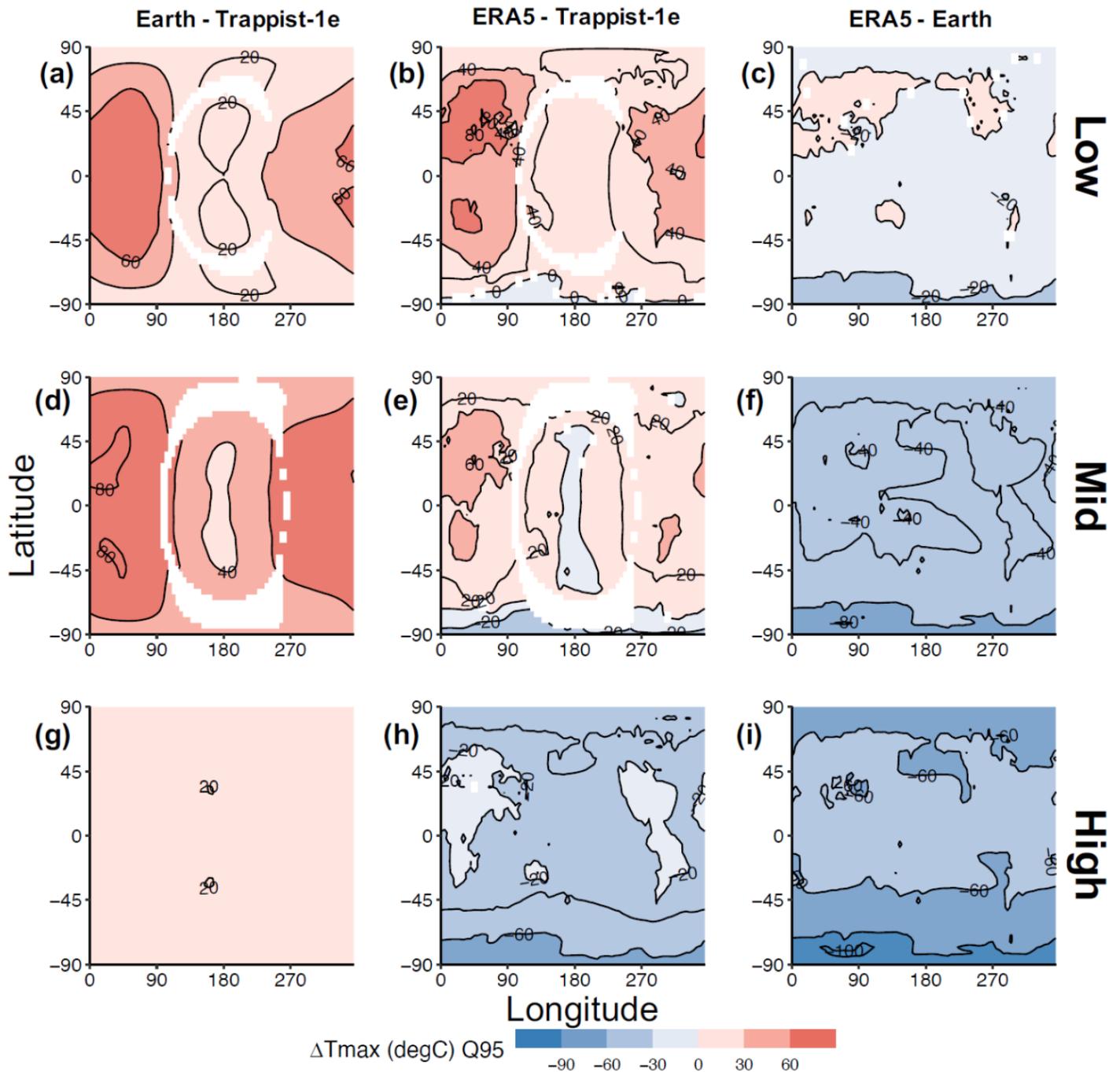

**Figure 5** Difference (Δ) between Earth-analogue and TRAPPIST-1e (a, d, g) spatial extremes for maximum temperature at the surface (*Tmax* in °C). (b, e, h) are the same as (a, d, g) but for ERA5 reanalysis and TRAPPIST-1e, and (c, f, i) for ERA5 reanalysis and the Earth analogue. The different pCO₂ scenarios are described in Section 3.1 and are referred to as 'Low' – $10^{-2}$ Bar, 'Mid' – $10^{-1}$ Bar and 'High' – 1 Bar. White regions represent areas without extremes and *non-statistically* significant values (p-value ≥ .01) following a two-tailed Wilcoxon rank sum test and a Bonferroni correction (see Sect. 3.4).

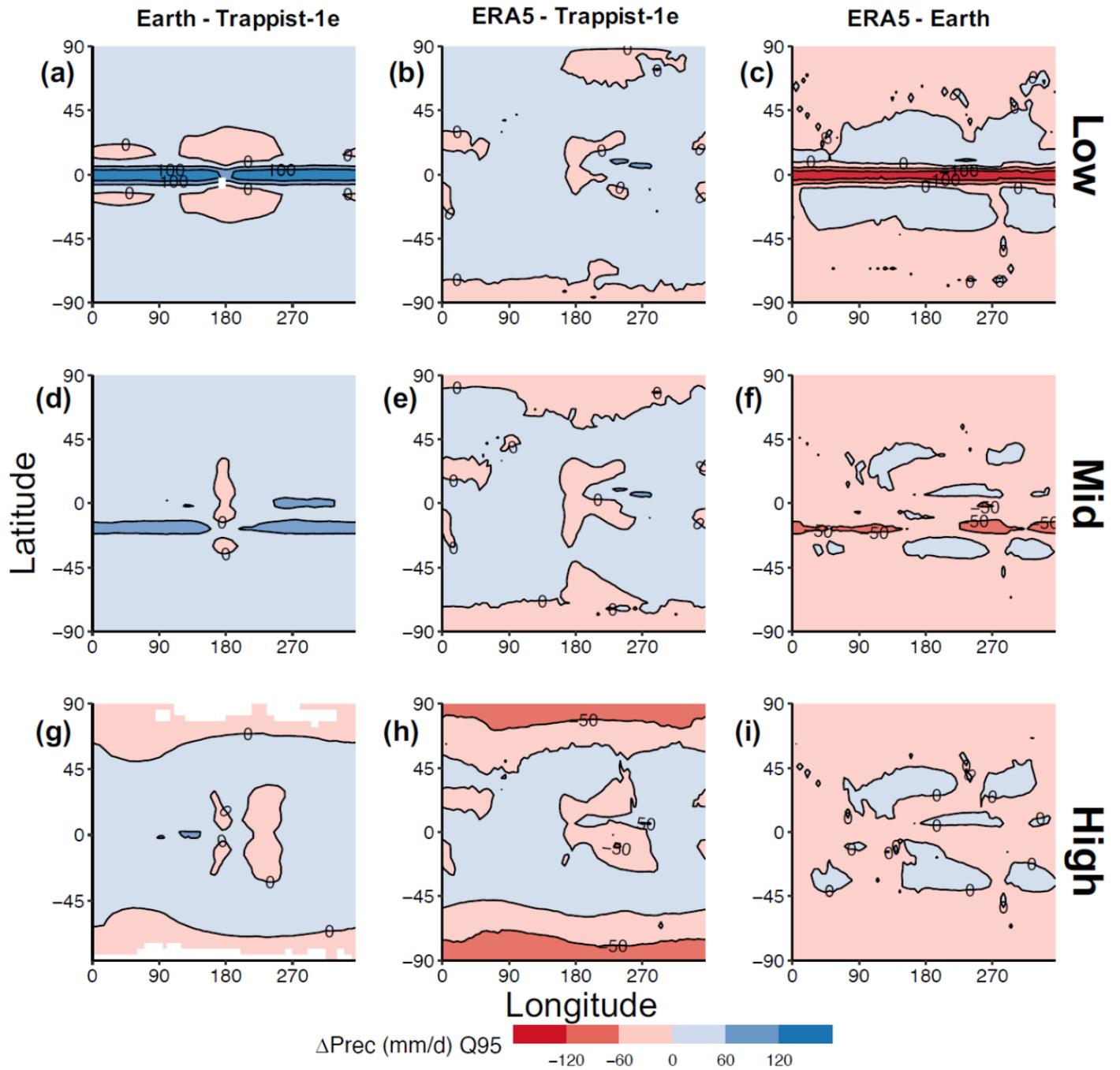

**Figure 6** Same as Fig. 5 but for precipitation (*Prec* in mm d⁻¹).

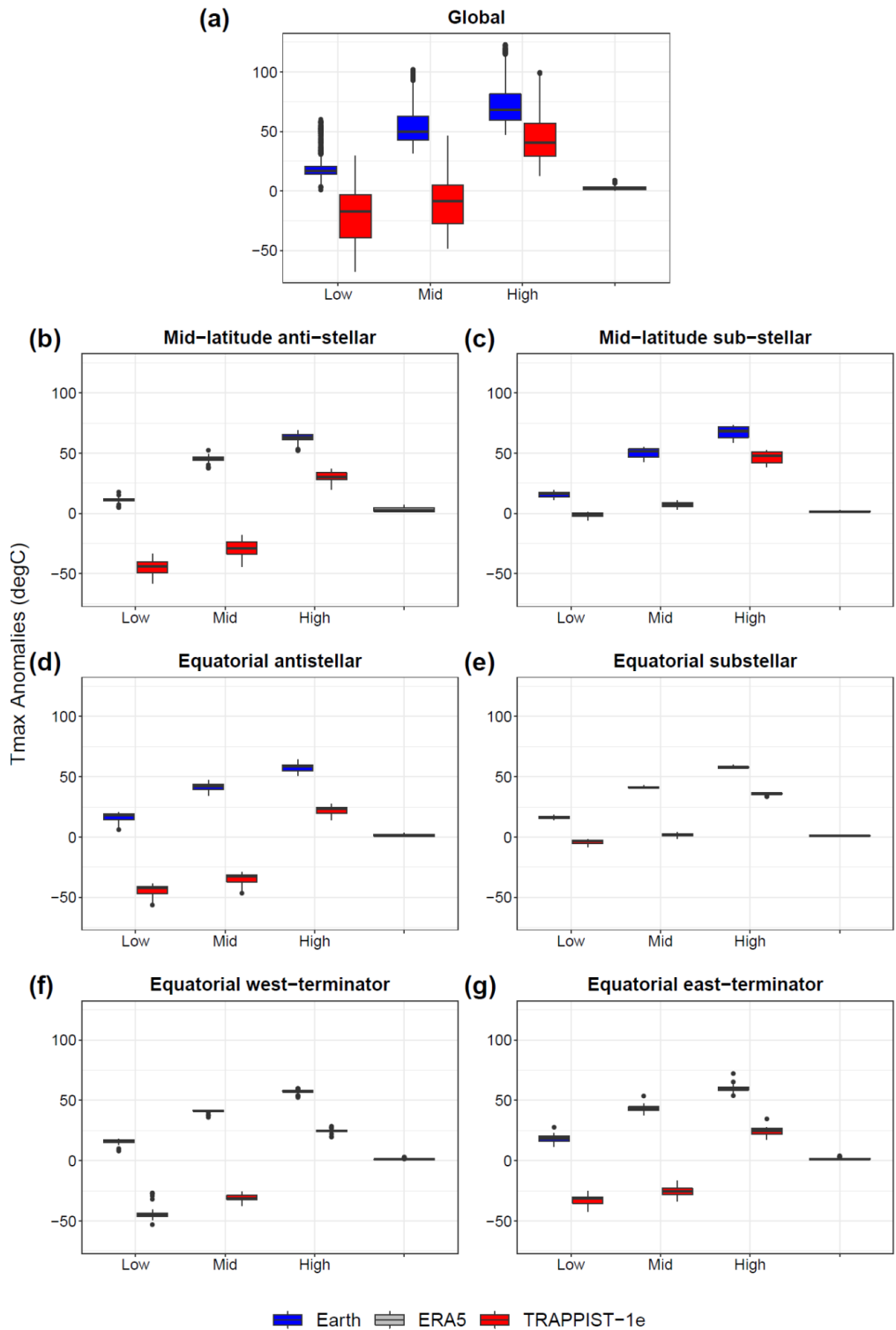

**Figure 7** Boxplots of daily maximum temperature (°C) of mean extreme anomalies computed with respect to ERA5 reanalysis climatology (1979-2020) for Earth (blue), TRAPPIST-1e (red) and ERA5 (grey). (b - g) Regions described in Fig. 1 a. The different $pCO_2$ scenarios are described in Section 3.1 and are referred to as 'Low' – $10^{-2}$ Bar, 'Mid' – $10^{-1}$ Bar and 'High' – 1 Bar. Extremes are computed for each grid-box as daily values exceeding the 95[th] percentile. The boxplots show the 25[th] quantile, median and 75[th] quantile. Black dots are outliers.

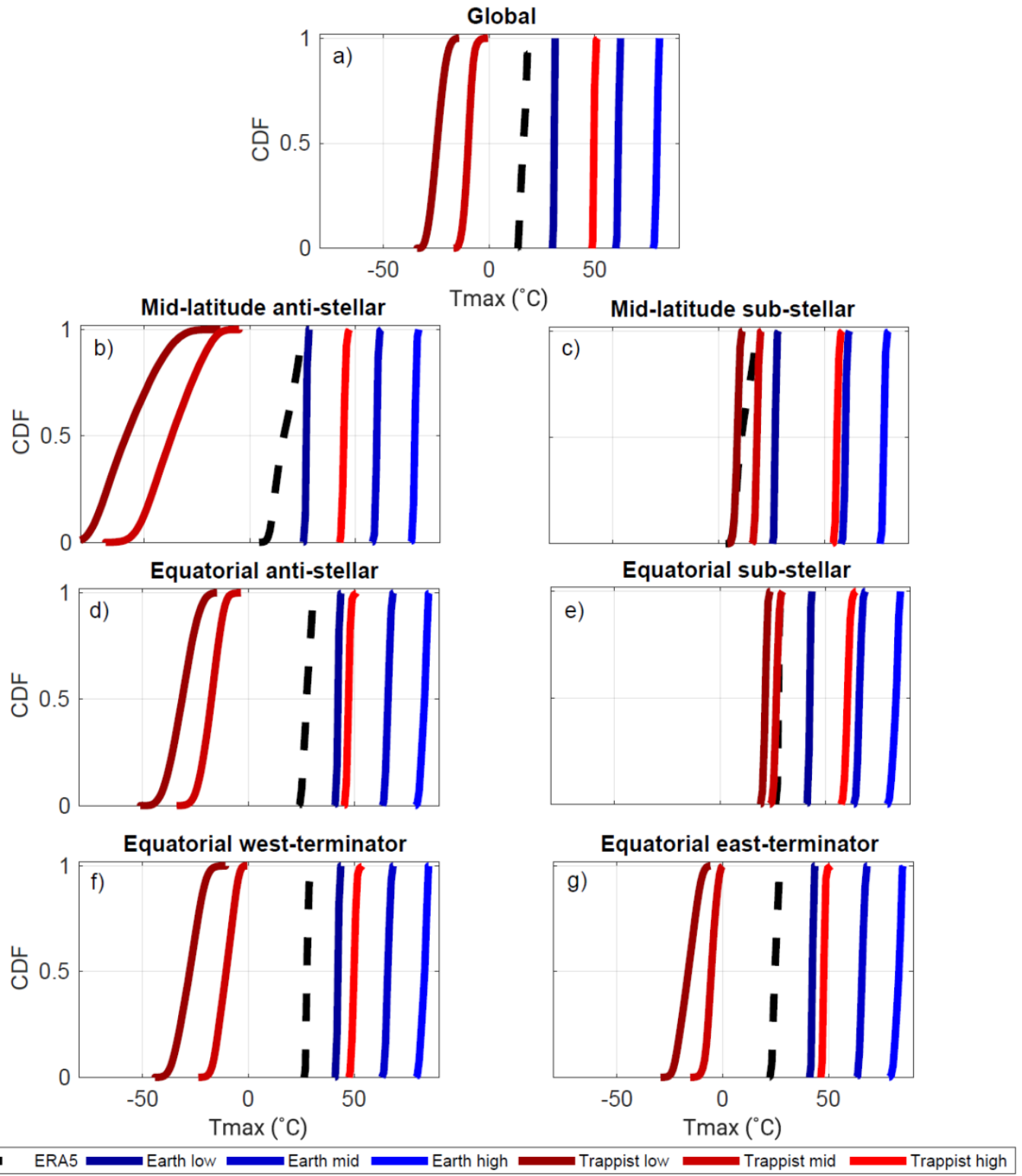

**Figure 8** Cumulative Distribution Functions (CDF) of median daily maximum temperature (*Tmax* in °C) for Earth (solid blue lines), TRAPPIST-1e (solid red lines) and ERA5 (dashed black lines). (b - g) Regions described in Fig. 1 a. The different pCO₂ scenarios are described in Section 3.1 and are referred to as 'Low' – $10^{-2}$ Bar, 'Mid' – $10^{-1}$ Bar and 'High' – 1 Bar.

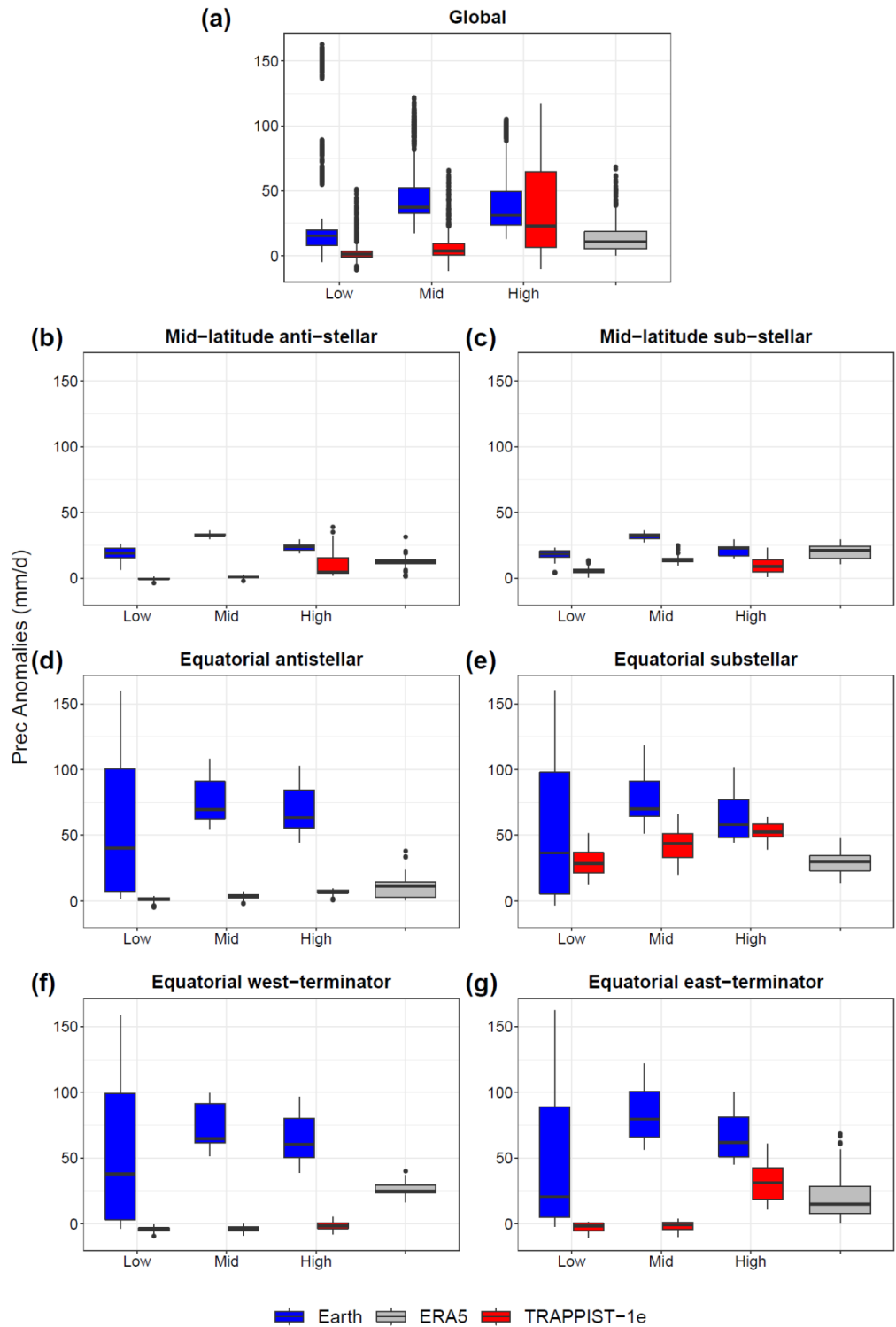

**Figure 9** Same as Fig. 7 but for daily total precipitation (*Prec* in mm d⁻¹).

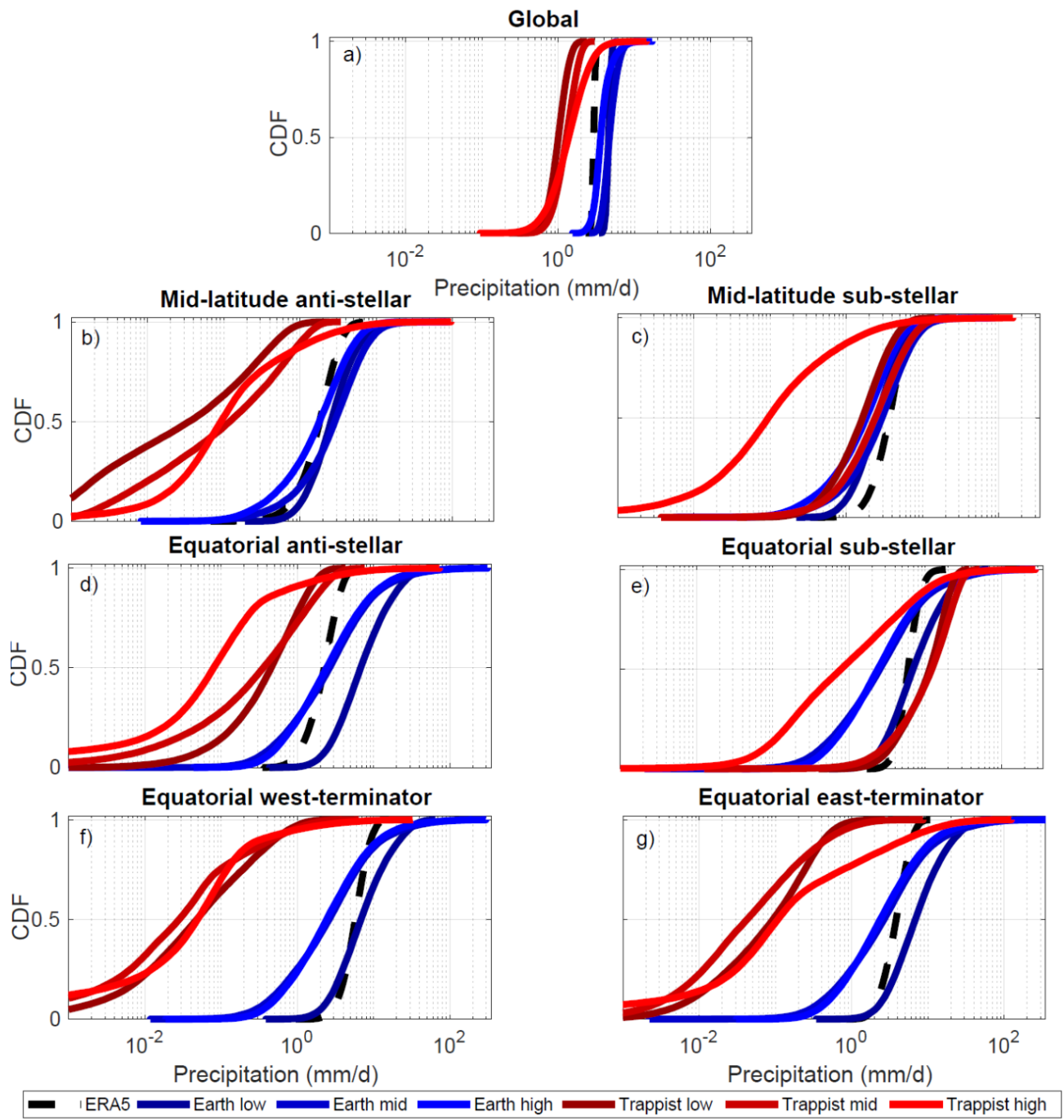

**Figure 10** Same as Fig. 8 but for daily total precipitation (*Prec* in mm d⁻¹).

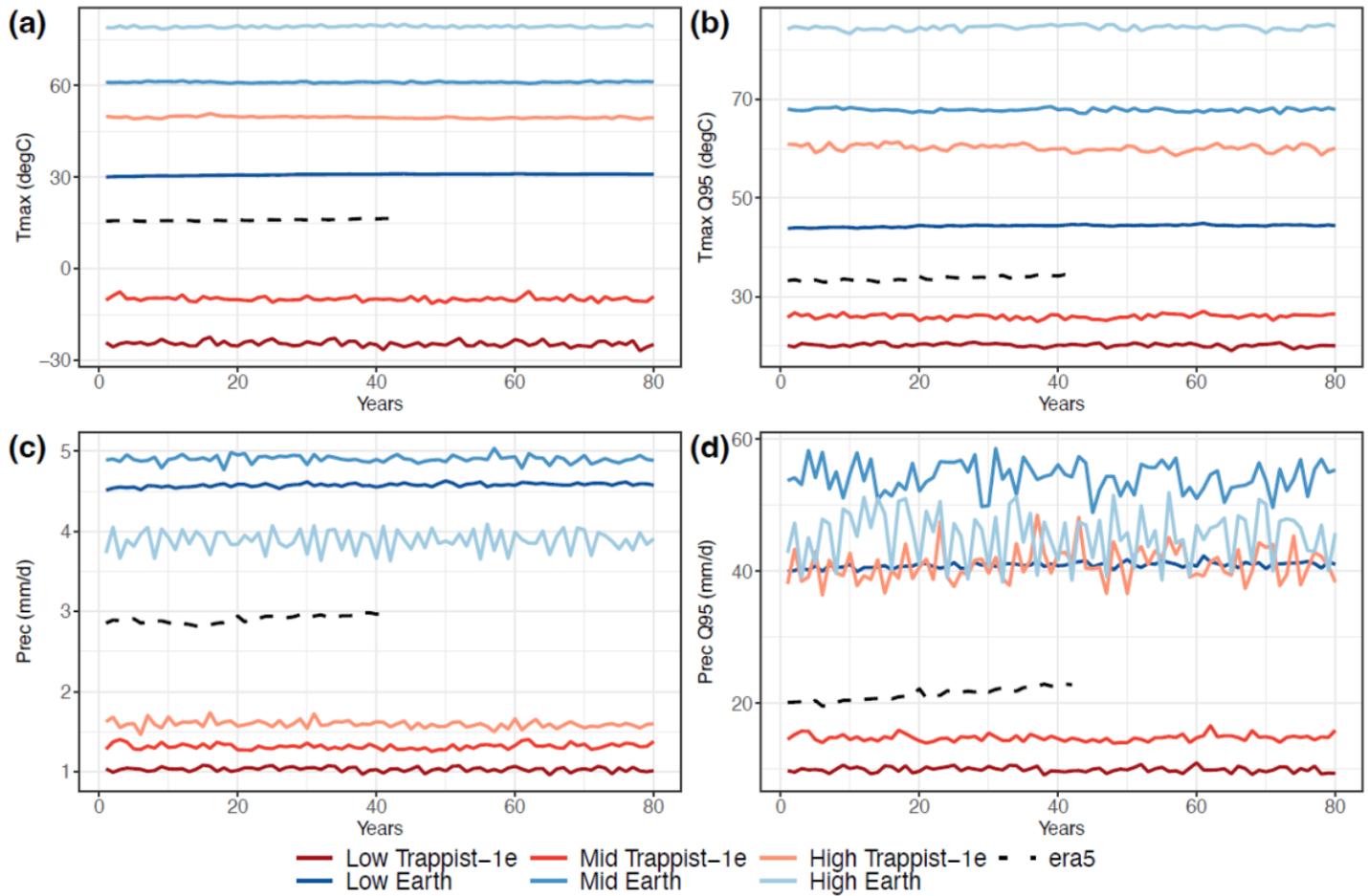

**Figure 11** Time series for annual global means and extreme (95th percentile - Q95) of daily maximum temperature (a, b; *Tmax* in °C) and precipitation (c, d; *Prec* in mm d$^{-1}$). Earth analogue (solid blue lines), TRAPPIST-1e (solid red lines) and ERA5 (dashed black lines). The different pCO$_2$ scenarios are described in Section 3.1 and are referred to as 'Low' – 10$^{-2}$ Bar, 'Mid' – 10$^{-1}$ Bar and 'High' – 1 Bar.

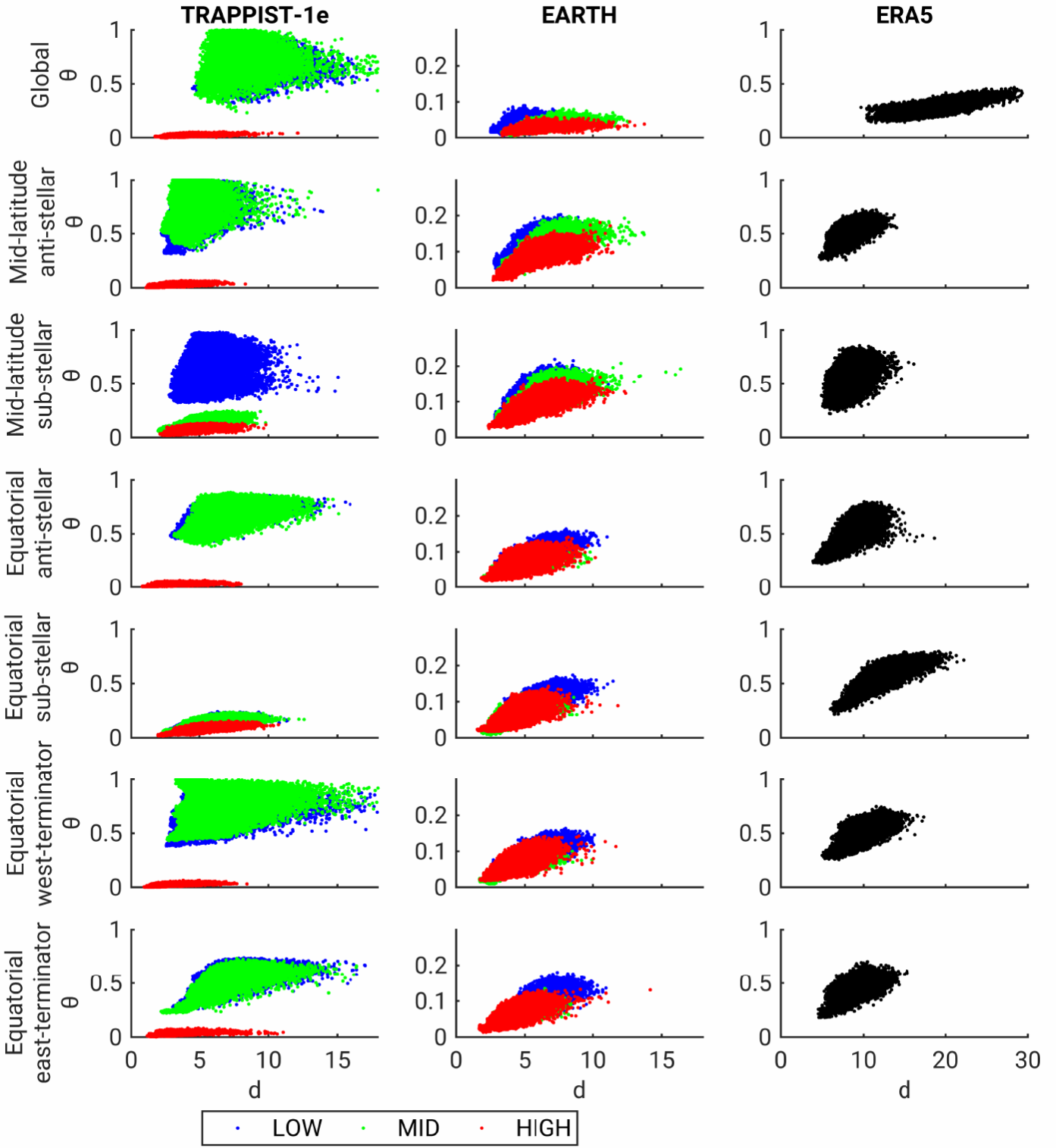

**Figure 12** Phase plane diagrams of the dynamical systems metrics $d$ (local dimension) and $\theta$ (inverse persistence) with varying pCO$_2$ ('LOW' - blue, 'MID' - green, 'HIGH' - red; see Section 3.1) for TRAPPIST-1e (left column), EARTH (middle column, note reduced y-axis scale) and ERA5 reanalysis as reference (right column, note extended x-axis scale). The regions used to compute the dynamical systems metrics are displayed in Fig. 1 a. The dynamical systems metrics are computed for the *Tmax* variable.

**APPENDIX**

This section includes complementary analysis to the main part of the manuscript. Indeed, in the main text we focus on the maximum temperature (*Tmax*) and precipitation variables, whereas here we present the same analysis as in the main text, but rather for minimum temperature (*Tmin*). We include this analysis in the Appendix because there are no significant differences in our interpretation of the results for *Tmin* as compared to *Tmax*. The Appendix includes Figures (A1 – A7), which are the same as Figures (1, 3, 5, 7, 8, 11 and 12), respectively, but for *Tmin* rather than *Tmax*.

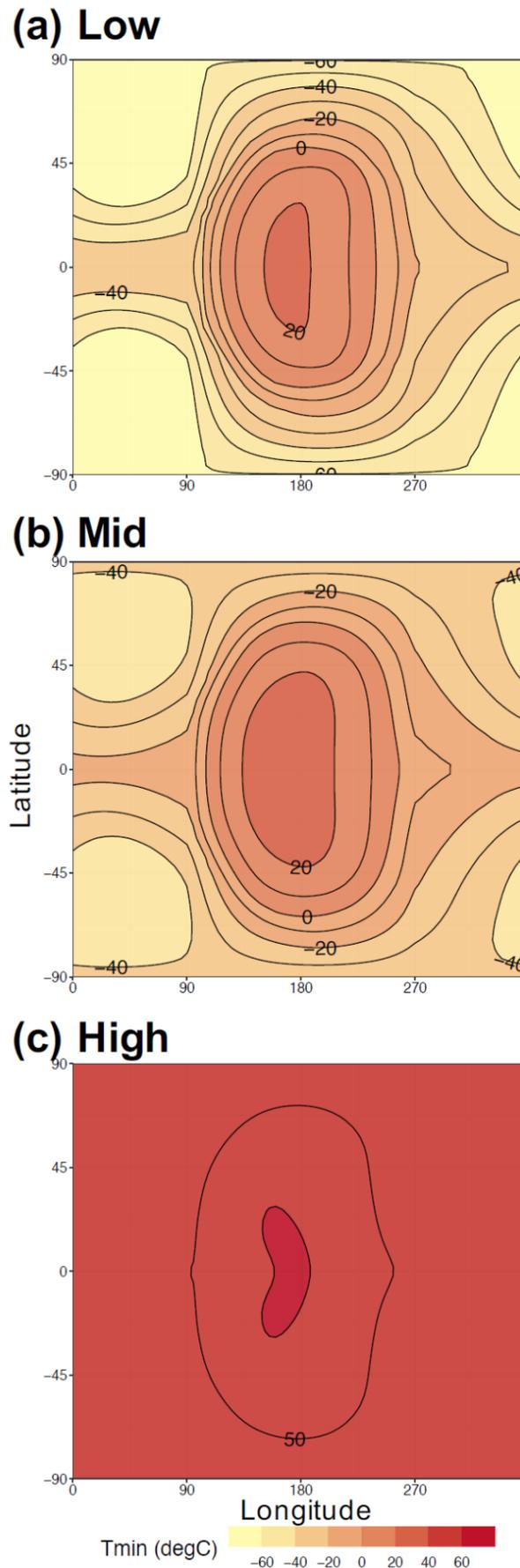

**Figure A1** Same as Fig. 1 but for minimum temperature (*Tmin* in °C).

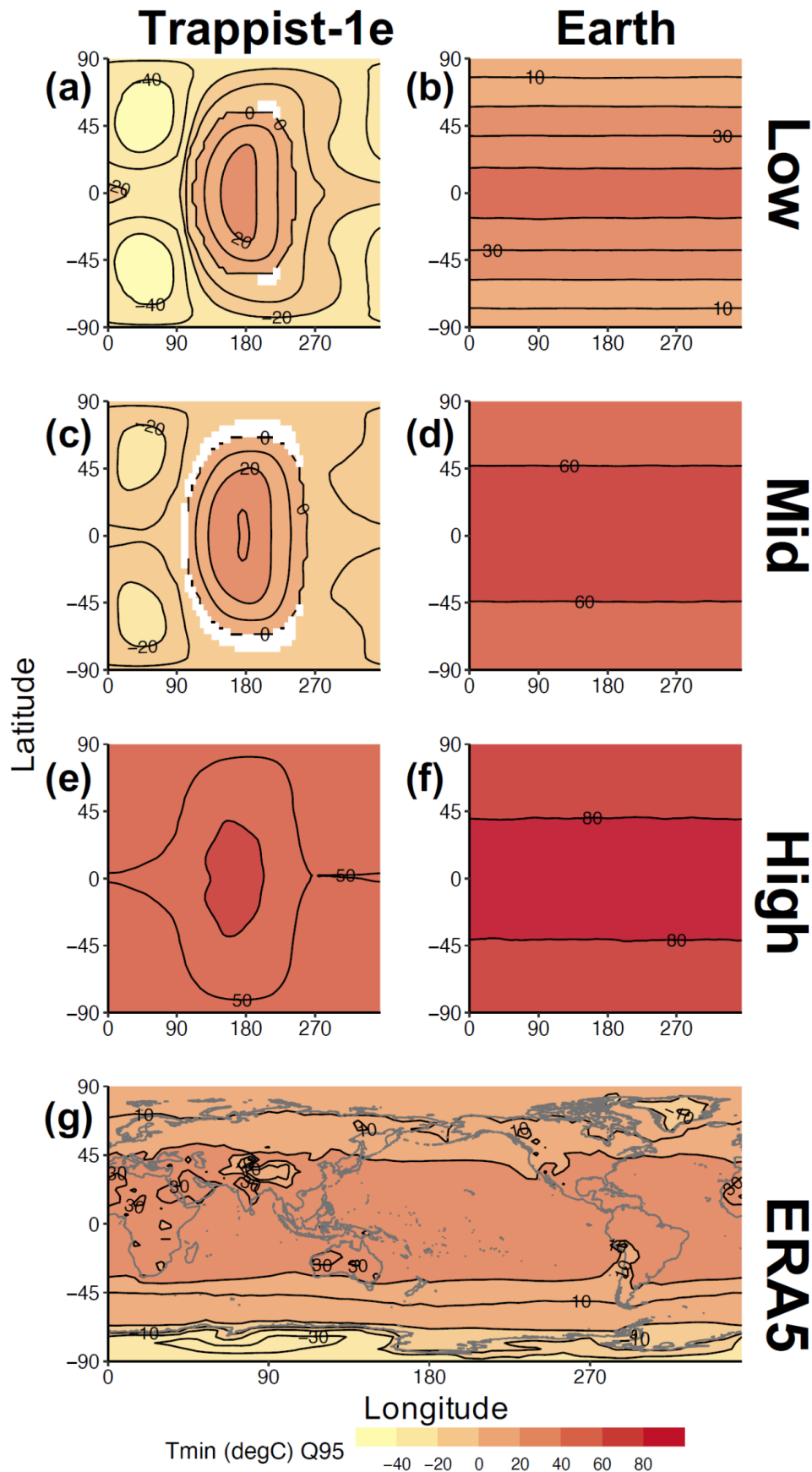

**Figure A2** Same as Fig. 3 but for minimum temperature (*Tmin* in °C).

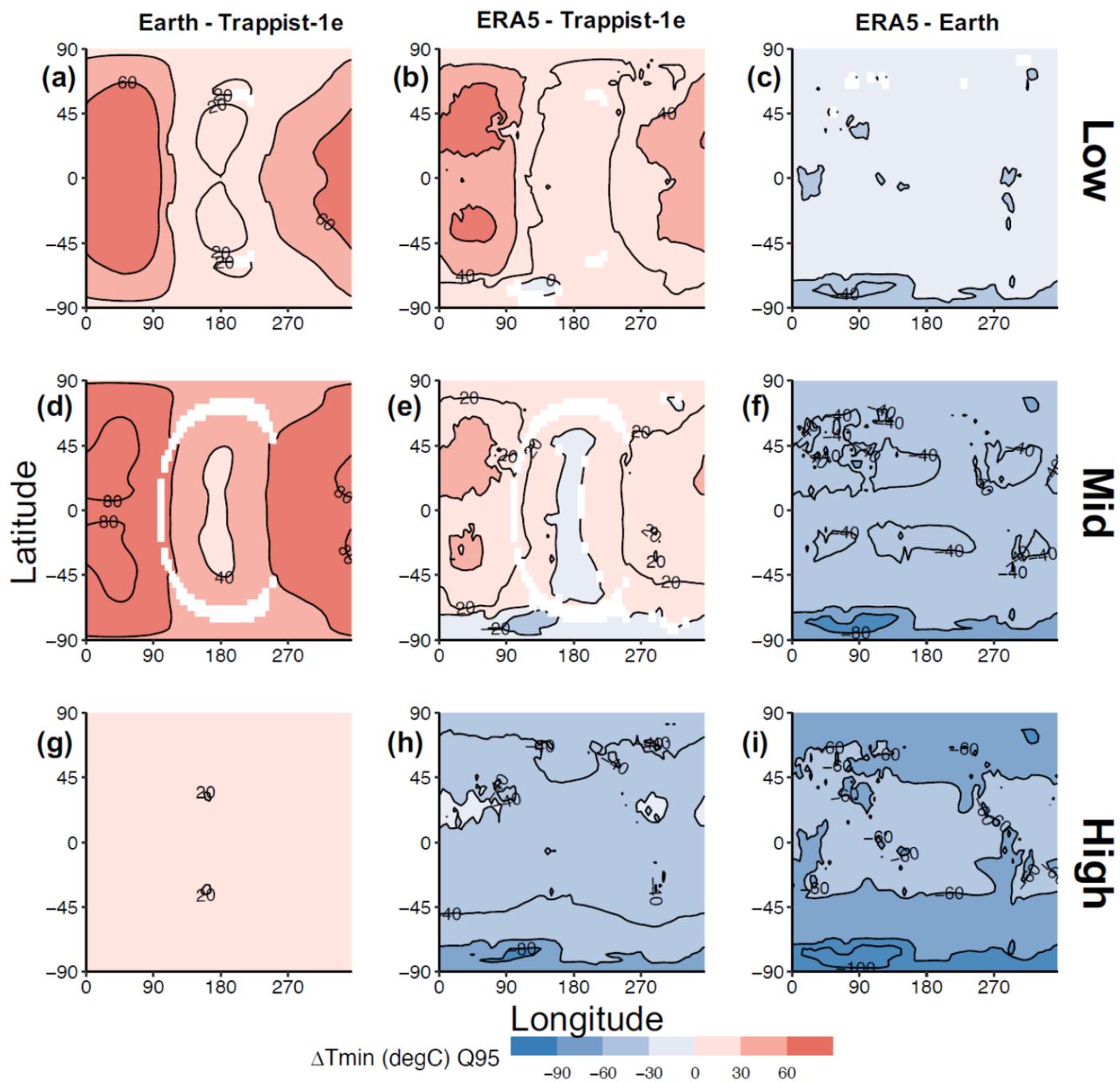

**Figure A3** Same as Fig. 5 but for minimum temperature (*Tmin* in °C).

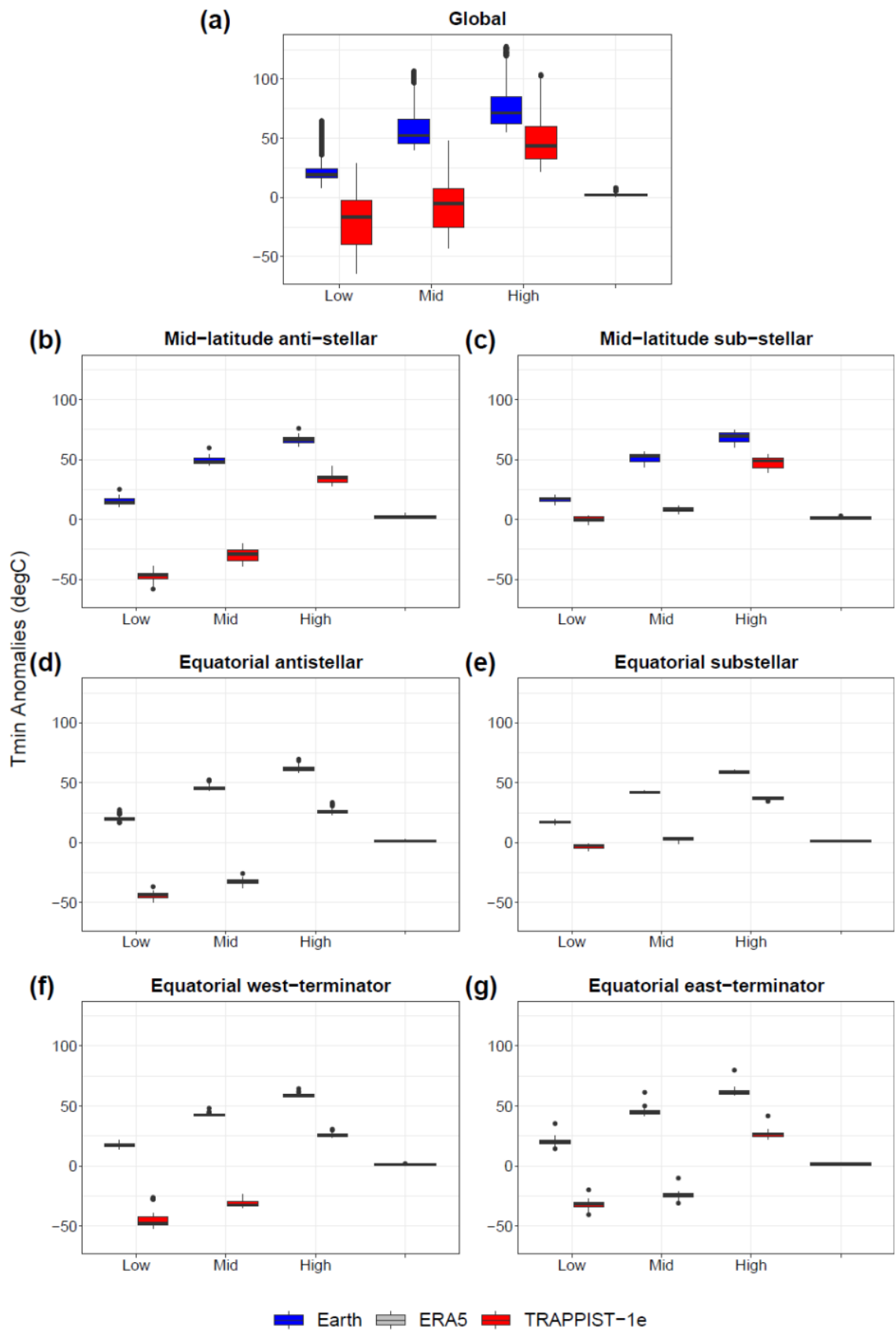

**Figure A4** Same as Fig. 7 but for minimum temperature (*Tmin* in °C).

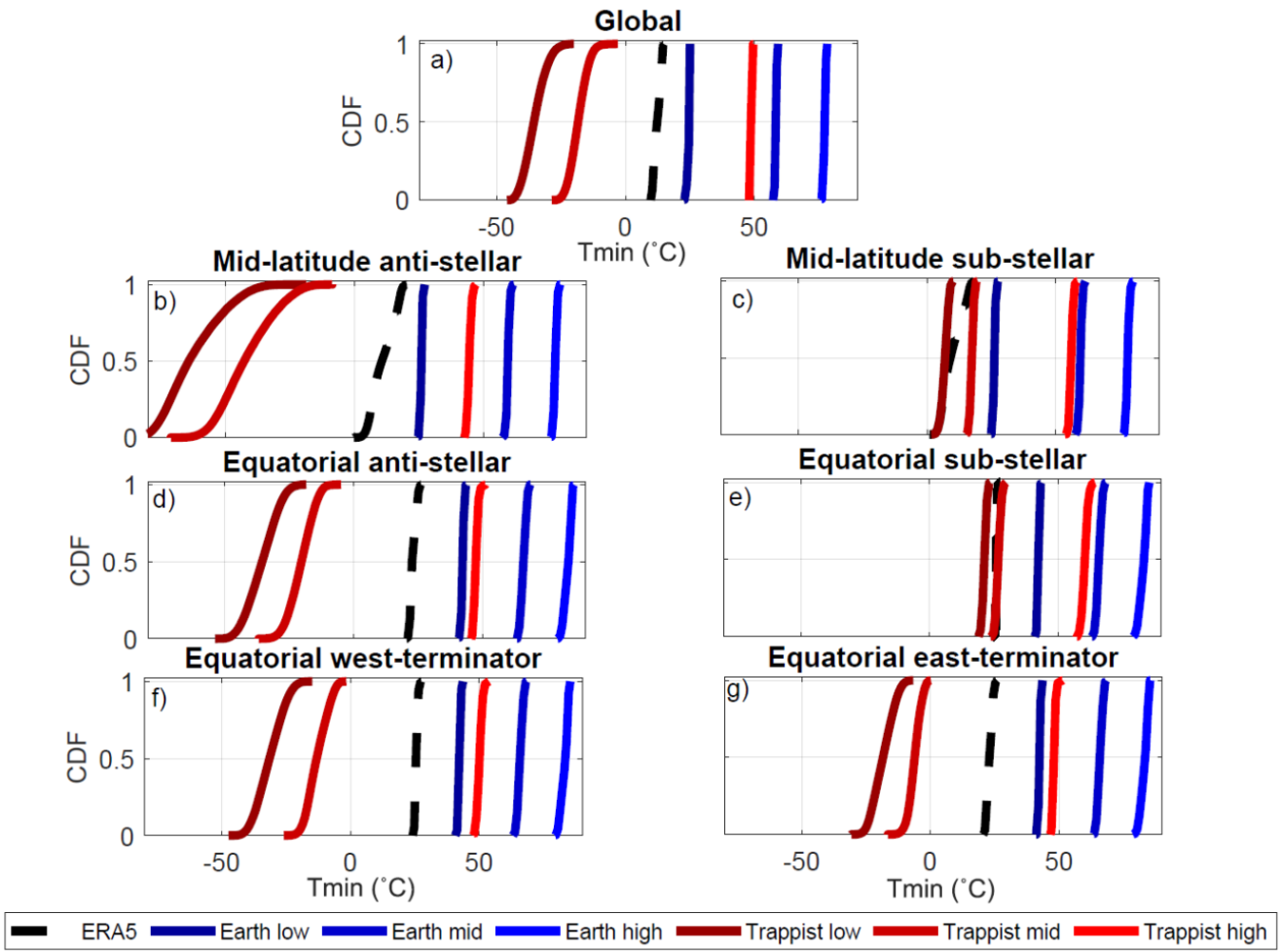

**Figure A5** Same as Fig. 8 but for minimum temperature (*Tmin* in °C).

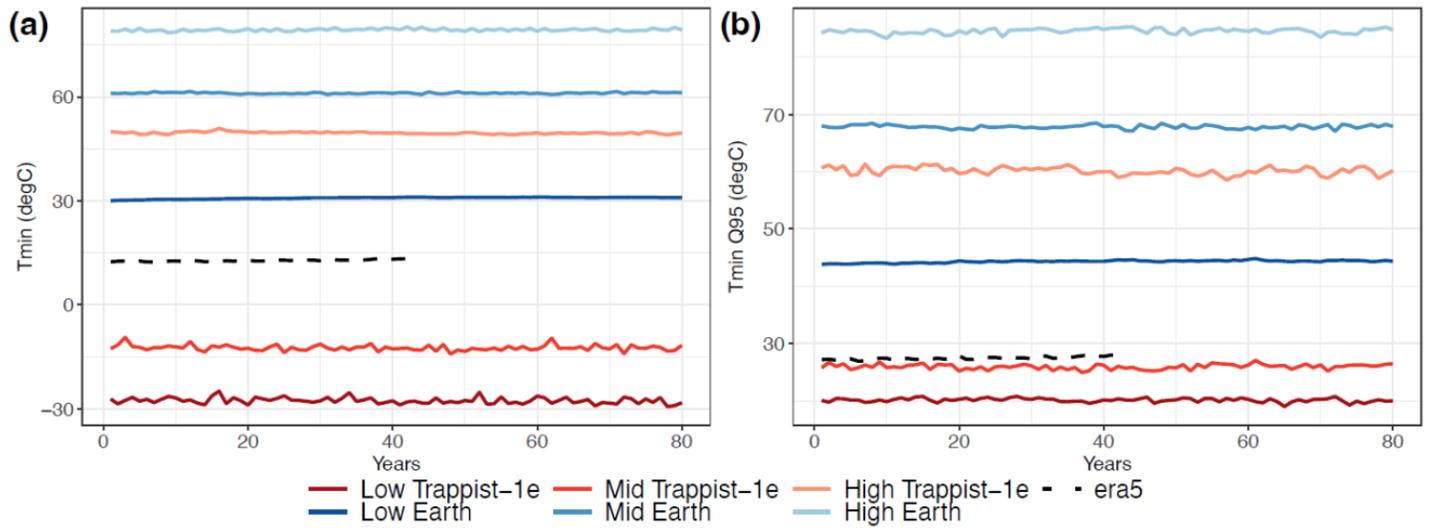

**Figure A6** Same as Fig. 11 but for minimum temperature (*Tmin* in °C).

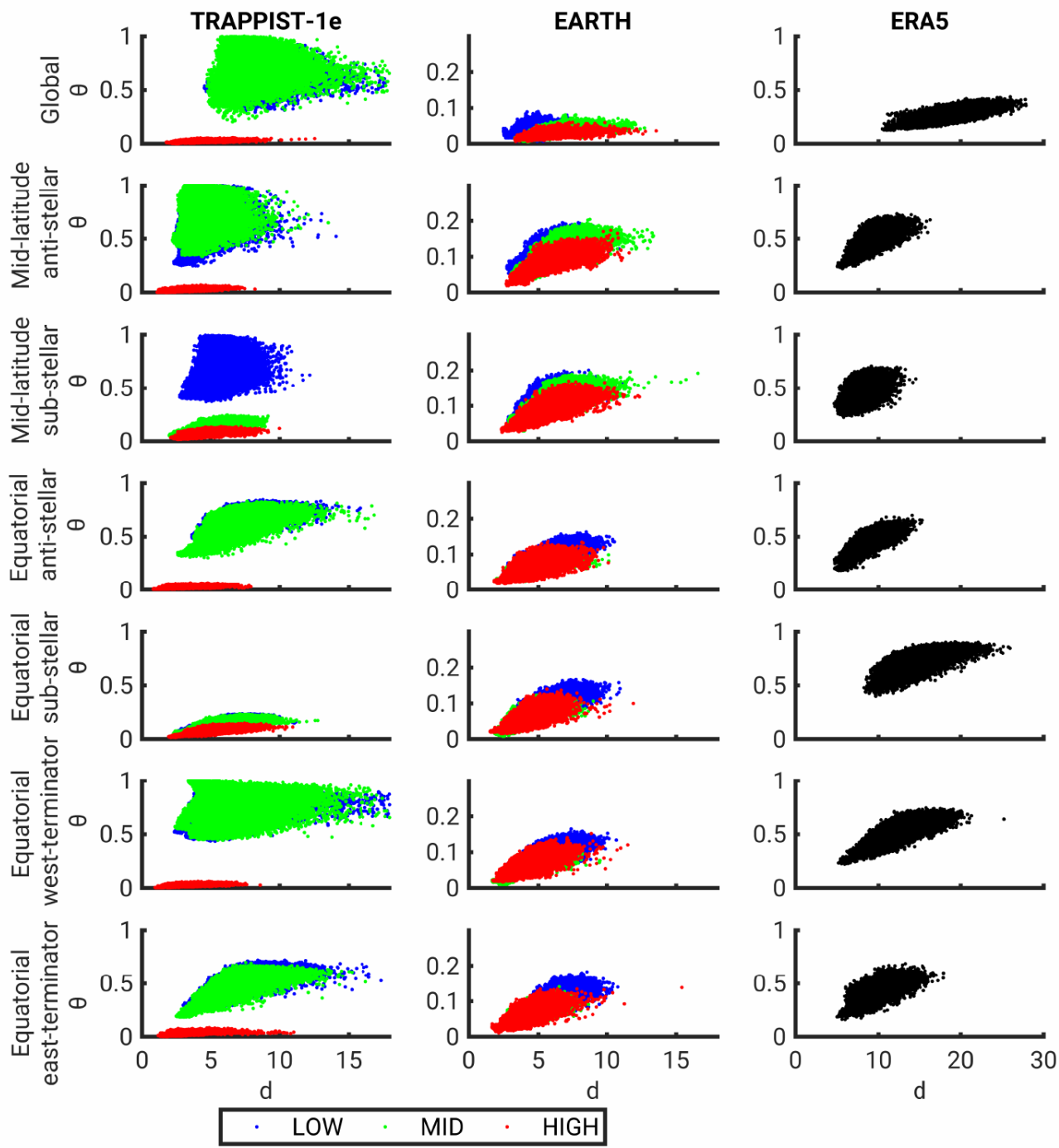

**Figure A7** Same as Fig. 12 but for the dynamical systems metrics computed on daily minimum temperature (*Tmin* in °C).